\documentclass[manuscript]{aastex}
\makeatletter
\let\@dates\relax
\makeatother
\usepackage{subfigure}
\usepackage{natbib}
\usepackage{graphicx}
\usepackage{multirow}
\usepackage{epstopdf}
\usepackage{amsmath}
\usepackage{url}
\usepackage{hyperref}
\usepackage{amssymb}
\usepackage{booktabs}
\usepackage{lineno}
\usepackage{color}
\shorttitle{GBM Updates for LIGO O2a}
\shortauthors{Goldstein, et al.}

\begin{document}

%%Title
\title{Updates to the {\it Fermi}-GBM Short GRB Targeted Offline Search in Preparation for LIGO's Second Observing Run}

%%Authors
\author{Adam~Goldstein\altaffilmark{1}, Eric~Burns\altaffilmark{2}, Rachel~Hamburg\altaffilmark{3}, 
Valerie~Connaughton\altaffilmark{1}, P$\acute{\rm e}$ter Veres\altaffilmark{3}, Michael~S.~Briggs\altaffilmark{3}, 
C.~Michelle~Hui\altaffilmark{3}, \& The~GBM--LIGO~Collaboration}
\altaffiltext{1}{Science and Technology Institute, Universities Space Research Association, Huntsville, AL 35805, USA}
\altaffiltext{2}{Department of Physics, University of Alabama in Huntsville, 320 Sparkman Drive, Huntsville, AL 35899, USA}
\altaffiltext{3}{Center for Space Plasma and Aeronomic Research, University of Alabama in Huntsville, 320 Sparkman Drive, 
Huntsville, AL 35899, USA}
\altaffiltext{3}{Astrophysics Office, ZP12, NASA/Marshall Space Flight Center, Huntsville, AL 35812, USA}

%% Introduction
\section{Introduction}
The {\it Fermi} Gamma-ray Space Telescope's Gamma-ray Burst Monitor (GBM) is currently the most prolific detector of 
Gamma-ray Bursts (GRBs), including short-duration GRBs.  The GBM triggers on $\sim$ 240 GRBs per year, $\sim$40 of which 
are short GRBs, localizes GRBs to an accuracy of a few degrees, has a broad energy band (8 keV--40 MeV) at high spectral 
resolution for spectroscopy, and records data at high temporal resolution (down to $\sim$2 $\mu$s)~\citep{Meegan09}.  Recently 
the detection rate of short GRBs has been increased with a ground-based search to detect fainter events which did not trigger 
GBM\footnote{\url{http://gammaray.nsstc.nasa.gov/gbm/science/sgrb_search.html}}.  The detection of GRBs by GBM have led to a 
plethora of analyses, including joint analyses with the {\it Fermi} Large Area Telescope, {\it Swift}, and ground-based optical and 
radio telescopes.  Although the localization of GRBs by GBM is rough in comparison to the capabilities of {\it Swift}, recent 
developments~\citep{Connaughton15} have led to the ability of wide-field optical telescopes such as the Palomar Transient 
Factory to scan the large GBM localization regions and discover the GRB optical afterglow independent of any other 
instrument~\citep{Singer15}.

Motivated by the possibility that short GRBs are caused by compact binary mergers that produce gravitational 
waves and may be observable by LIGO, \citet[][hereafter LB15]{Blackburn15} developed a method to search the GBM continuous 
data for transient events in temporal coincidence with a LIGO compact binary coalescence trigger.  The LB15 search operates by 
ingesting a LIGO trigger time and optionally a LIGO localization probability map and searches for a signal over different 
timescales around the LIGO trigger time.  The search looks for a coherent signal in all 14 GBM detectors by using spectral 
templates which are convolved with the GBM detector responses calculated over the entire un-occulted sky to produce an 
expected count rate signal in each detector.  The expected counts in each detector are compared to the observed counts, taking 
into account a modeled background component.  A likelihood ratio is calculated comparing the presence of a signal to the null 
hypothesis of pure background, and the likelihood ratio is marginalized over the spatial prior.  If a LIGO sky map is provided, then 
this is used as the spatial prior, otherwise a uniform prior on the sky is used.  The log-likelihood ratio (LogLR) for a uniform sky 
prior and the spatially-coincident log-likelihood ratio (CoincLR) are compared between the templates, and the template with the 
largest LogLR or CoincLR is selected for consideration as a real signal.  LB15 discusses the formalism and implementation of this 
approach along with an estimation of the detection significance distribution during 2 months of the LIGO S6 run.

In preparation for the first Advanced LIGO Observing run (O1), the LB15 search was implemented into a GBM pipeline to search 
for gamma-ray signals coincident with LIGO triggers.  The first use of the pipeline was on September 16, 2015 to search for any 
candidate counterparts to what is now known to be the first direct observation of a binary black hole coalescence, 
GW150914~\citep{Abbott16}.  That search resulted in a low-significance, spectrally hard candidate at $\sim$0.4 s after the LIGO 
trigger time.  The analysis of the candidate, which is observationally consistent with a weak short GRB arriving at a poor geometry 
relative to the GBM detectors, is described in~\citet{Connaughton16}.  The pipeline was used three other times during O1, one of 
the runs was on a LIGO trigger that was later retracted, and the other two were run on the LIGO triggers for LVT151012 and 
GW151226.  As detailed in~\citet{Racusin16}, no significant gamma-ray candidates were found for either event.

As LIGO has undergone a hiatus from January 2016 until Fall 2016 to improve and upgrade the instrument, the GBM team has 
also tested and implemented updates and improvements to the original LB15 pipeline.  These updates, discussed in the following 
sections, will be used during the LIGO O2a run to search for coincident gamma-ray counterparts in the GBM data.  Areas for future 
improvement are noted, and further updates are planned for upcoming observing runs.

%% Data resolution
\section{Input Data\label{sec:InputData}}
The LB15 search was designed to use the continuous CTIME data, which has a 256 ms temporal resolution and 8 energy 
channels.  In November 2012, GBM began producing continuous Time-Tagged Event (TTE) data, which contains the time and 
energy channel information for each individual incident photon, down to 2 $\mu$s resolution.  The continuous TTE is now 
downlinked from the spacecraft every couple of hours, while the continuous CTIME is assembled on a daily cadence.  The need 
for rapid reporting in the event of a LIGO trigger is better satisfied by using the continuous TTE, so we have implemented code to 
bin the continuous TTE as it arrives for use with the offline search.  Although the TTE data is produced at 128-channel 
energy resolution, we currently choose to combine events into the same 8 energy channels defined by the CTIME data.  We have 
chosen to change the temporal resolution of the binning from the CTIME-native 256 ms to a 64 ms binning, which allows future 
searches at shorter timescales and with more flexibility for phase shifting in the search.  We currently continue to execute the 
search over timescales from 256 ms to 8.192 s in factors of 2, but with a phase shift of a factor of 4 when using the 64 ms binning.  
At this time, we also continue to operate the search with the maximum phase shift factor of 4 imposed during O1 (i.e. 256 ms 
phase shift for the 1024 ms timescale even though the bin resolution is 64 ms).

%% Background Estimation
\section{Background Estimation\label{sec:NewBack}}
The LB15 search fit the background with a ($\geq 2$) polynomial to model the typically smoothly-varying background 
around a time of interest.  A short window on either side of the candidate time was removed from background consideration, and 
then a window of duration from $\sim10-100$ s was used on either side of this `dead' period.  The duration of the data used for the 
background fitting depended on the time integration of the candidate, with the shorter background window applied to 0.256 s 
duration candidates, and $\sim100$ s background window applied to 8.192 s candidates.  The background was fit using a 
least-squares minimization routine, and the variance of the polynomial coefficients and covariance matrix was used to estimate the 
variance of the fitted background during the candidate time.  The variance from the fit was combined with the Poisson background 
rate variance for use in the LB15 likelihood calculation.  Using this method, the background in each of the 8 CTIME channels for 
each detector was estimated.  Note that typically the lowest number of counts in an individual channel and time bin is $> 10$, 
therefore a Gaussian treatment of the background variance from the fit is a reasonable approximation to the full Poisson 
calculation.

We have updated the background estimation to an un-binned Poisson maximum likelihood technique using the continuous TTE 
data.  For a background that is generated by a Poisson process defined by a single, time-independent rate, the un-binned 
likelihood can be written as
\begin{equation}
L = \prod_{i=0}^M e^{-t \lambda} \prod_{j=0}^N t \lambda e^{-t \lambda},
\label{eq:PoissonLike}
\end{equation}
where the first product represents the $M$ `bins' that contain no count, and the second product represents the $N$ `bins' that 
contain a single count.  The `bin width', $t$, is set by the minimum allowed time between events, therefore requiring that no bin 
can have more than one count.  The value of the rate, $\lambda$, that maximizes the likelihood can be found by solving for a zero 
first derivative with respect to the rate.  Because the logarithm of the likelihood preserves the location of the maximum, typically 
the derivative is taken on the log-likelihood.  The $\lambda$ maximum likelihood solution for Equation~\ref{eq:PoissonLike} is
\begin{equation}
\lambda_{\rm max} = \frac{N}{(N+M) t}.
\label{eq:MaxLike}
\end{equation}
This solution is simply the ratio of the number of `bins' with a count to the total number of `bins' in the considered data.  While this 
solution is used to estimate a constant Poisson rate, we can adapt this to use with time-varying GBM data.  If we define a 
large time window of duration $T$, over which Equation~\ref{eq:MaxLike} is applied, and assuming that in this window the rate can 
be approximately described by a constant Poisson rate, then we can slide this window through the data, at each point estimating 
the Poisson rate.  This approach can fail where there exists either a relatively long and/or bright signal in the window, relative to 
the duration of the window; the estimated background will then be biased by the significantly increased rate due to the source.  In 
the presence of a short and/or weak signal in the time window, the bias is negligible, therefore this method is applicable for use 
with the LB15 search.  There are a number of possible ways of identifying regions of poor background estimation, and we discuss 
our implementation in Section~\ref{sec:BadBack}.

\begin{figure}
	\begin{center}
		\includegraphics[scale=0.8]{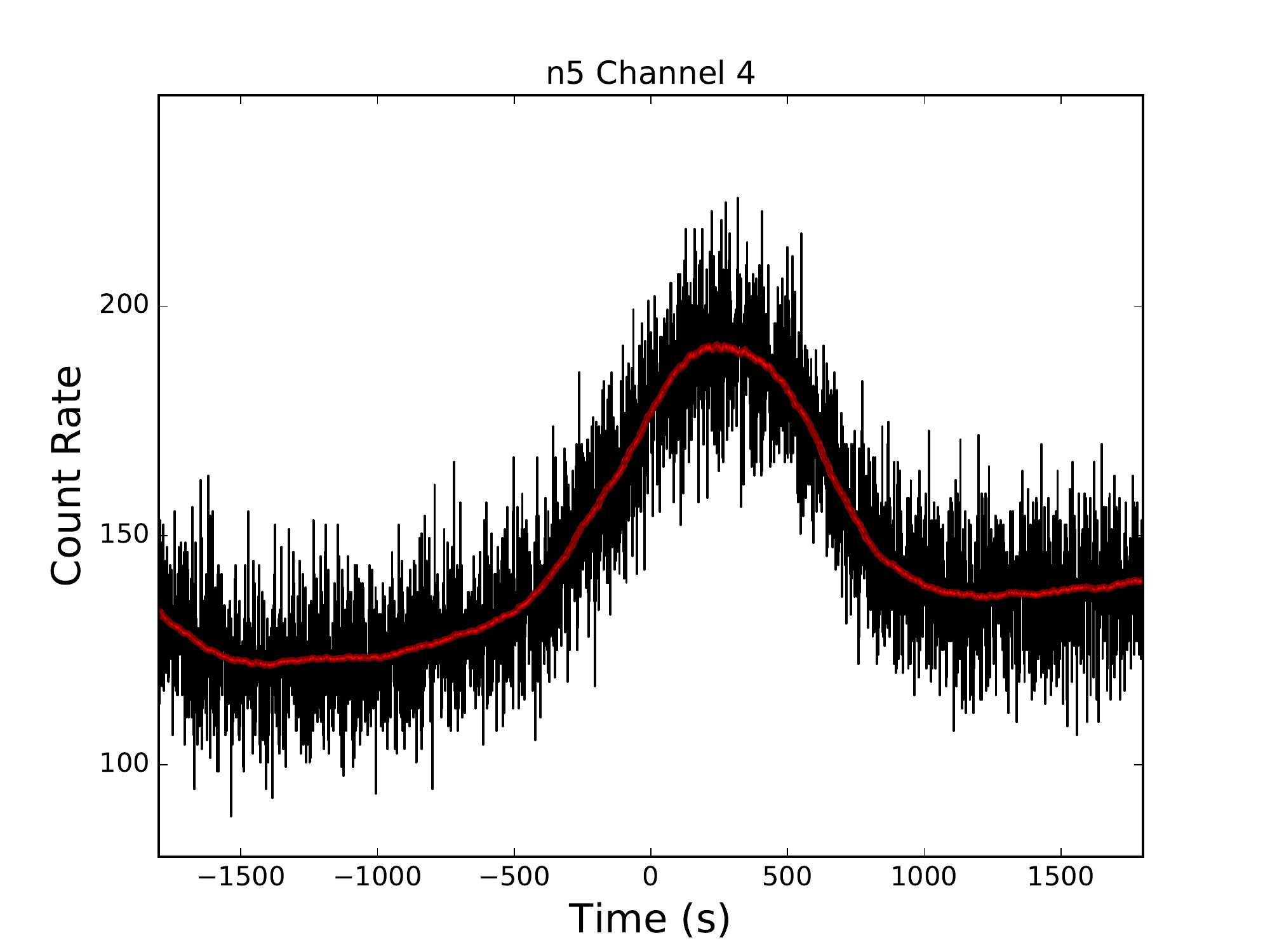}
	\end{center}
\caption{Background estimation of GBM data for a duration of an hour for CTIME channel 4 in NaI detector 5.
\label{HourBackground}}
\end{figure}

Applying this technique specifically to GBM data, the minimum time separation between two detected events allowed by the GBM 
onboard electronics is $2.6\ \mu$s, therefore any bin width $t$ shorter than this can be used.  The actual implementation of this 
algorithm, for purposes of speed and efficiency, allows the duration of the sliding window to vary by keeping the number of events 
in the window constant.  An additional benefit of allowing a varying window size is that the background can be properly estimated 
as the background rapidly (but smoothly) increases and decreases during periods of high particle activity and during entrance 
to/exit from the South Atlantic Anomaly (SAA).  Testing with several days of GBM data, an appropriate initial sliding window 
duration of $\sim$125 s for the average GBM background has been chosen.  Within the sliding window, Equation~\ref{eq:MaxLike} 
is calculated and  estimated as the background rate at the center of the window, thereby allowing the background rate to be 
estimated at the time of each individual TTE event.  In practice, this is likely overkill, and the rate can be down-sampled or 
interpolated to a regular interval.  An example of this background estimation is shown in Figure~\ref{HourBackground} for an hour 
of data.  Because the rate estimate is applied to the center of the window, the first half of the initial window and the second half of 
the final window will not have a background estimate.  We can, however, allow the sliding window duration to truncate at the 
beginning and end of these windows, which produces rate estimates with representatively larger uncertainties.  This method is 
particularly applicable when {\it Fermi} enters and exits the SAA.  The variance in the background rate estimate is simply $
\lambda/T$.  We calculate this uncertainty and include it in the likelihood calculation described in LB15.

%
%The new method, as described, can be implemented independent of the LB15 search, and in fact can be used to run continuous 
%background estimations for an indefinite period of time.  A test run of an implementation in Python on a single iMAC i5 CPU core 
%was able to process 8 continuous days of data from all 14 GBM detectors in $\sim$3.5 hours.
%
%% Template Spectra
\section{Template Spectra\label{sec:Templates}}
The LB15 search used a set of three template photon spectra, which are folded through each of the GBM detector responses and 
compared to the observed count spectra of a time of interest.  These spectra were the same three that are used in the GBM 
localization algorithms for GRBs produced by the GBM team~\citep[see][for details]{Connaughton15}, which has localized 
$>$1900 GRBs.  The spectra are all Band functions~\citep{Band93}, which is a smoothly-broken power law defined by low- and 
high-energy power-law indices, and the peak of the $\nu F_\nu$ spectrum, $E_{\rm peak}$.  The three template spectra are 
typically referred to as `Soft,' `Normal,' and `Hard' since the parameters were chosen to roughly and broadly sample the full 
spectral diversity of GRBs in the GBM energy band, from spectrally softer long GRBs to harder short GRBs.  In practice, the `Soft' 
spectrum is also used to localize SGR outbursts~\citep{van der Horst10} and other soft Galactic sources~\citep[e.g.][]{Jenke16}, of 
which many are present in the GBM data.  Although the spectral templates may not individually be a perfect match for any 
particular GRB (or GRB-like) spectrum, they span the range of observed spectral behavior of GRBs in the GBM band, and they 
have performed well in discovering particular weak signals in GBM data, as shown by the LB15 search.

While the soft and normal spectra are typically good matches for long GRBs, the hard spectrum has a very shallow (and 
unphysical) high-energy power-law index of -1.5.  Of more concern, however, is that very few GBM-triggered short 
GRBs are actually best fit by a Band function.  Instead, most detected short GRBs have a spectrum that is best-modeled as a 
power-law with an exponential high-energy cutoff:
\begin{equation}
	f(E) = A \ \biggl(\frac{E}{E_{\rm piv}}\biggr) ^{\alpha} \exp \biggl[ -\frac{(\alpha+2) \ E}{E_{\rm peak}}  \biggr].
\end{equation}
The maximum of the joint distribution for the cutoff power law for observed short GRBs is index, $E_{\rm peak}$~=~$\sim$
(-0.4,~600~keV), however some of the softer GRBs are adequately matched by the normal spectrum, so we choose to create a 
cutoff power law template with index, $E_{\rm peak}$~=~(-0.5,~1.5~MeV) which should better match the averaged spectrum of 
hard short GRBs.  We have chosen to replace the hard template used in LB15 with this new Comptonized template for O2.

%% Candidate Filters
\section{Candidate Filters}
The LB15 search was formulated and optimized to find weak short GRB-like events in the GBM data.  GBM data can be very 
complex and filled with many persistent and transient sources from all over the sky (and Earth), therefore it should be of no 
surprise that there exists phenomena uninteresting to the search that are nevertheless `discovered' by the search.  These 
phenomena contribute to the false alarm rate (FAR), which is an empirical distribution based on what the search finds.  The FAR 
estimation does not require perfect statistical technique, nor does it require efficiency in the search algorithm, however the FAR 
will be empirically lower for a statistically sound and efficient search compared to an inefficient search.  Specifically, efficient 
identification and rejection of signals that are obviously unrelated to GW sources will reduce the empirical FAR and thereby 
improve the significance of a true gamma-ray signal detected in conjunction with a GW signal. 

The LB15 search employed during the LIGO O1 run included a filter that removed single-detector phosphorescence 
events~\citep{Fishman77} from contaminating the FAR.  The FAR estimate over the two months of LIGO S6 run and for 2.7 days 
around the GW150914-GBM candidate included only this filter.  Inspection of some of the more significant false alarms found that 
they were caused by areas of poor background estimation, approach and exit from the SAA, solar flare events, and Galactic 
sources, and therefore we would like to filter out these types of false alarm events.

\subsection{Removing Bad Background\label{sec:BadBack}}
\begin{figure}
	\begin{center}
		\includegraphics[scale=0.8]{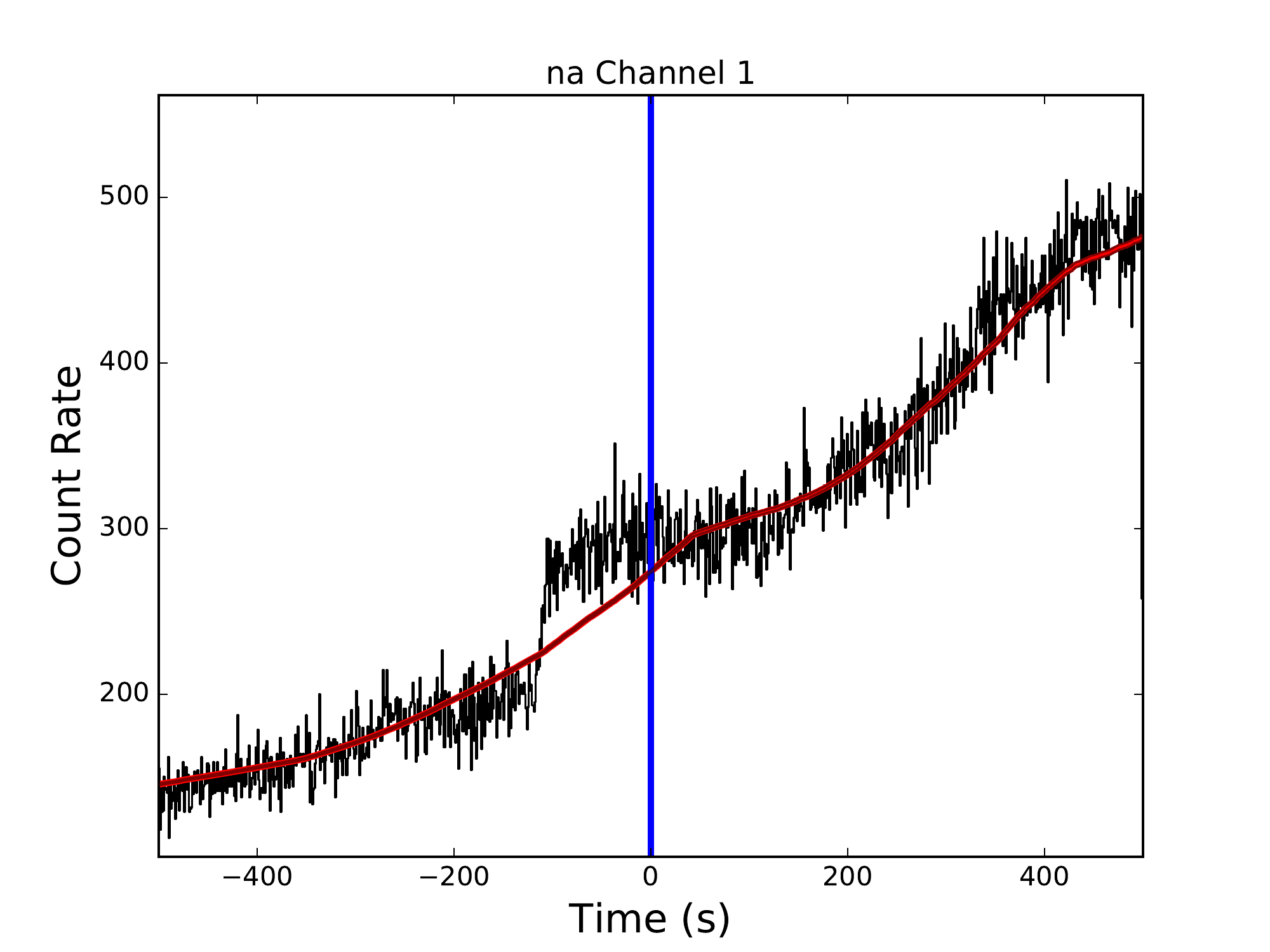}
	\end{center}
\caption{Background estimation during a long, soft source at $\sim$100 s before the time of interest for 
CTIME channel 1 in NaI detector 10 (also known as `NaI a').  The background estimate at the time of interest was flagged as bad 
due to the poor reduced chi-square statistic in the period prior to the candidate time.
\label{BadBackground}}
\end{figure}

We wish to filter out events due to bad background estimation.  In particular, the background estimation described in Section 2 can 
be biased by strong or very long sources in the data, as well as occultation steps in the lower energy channels when a bright 
persistent source rises or sets behind the Earth limb.  To determine if the background estimate is a poor representation of the true 
background at the time of interest, we consider a `dead' period of 10 s on either side of the time of interest.  Beyond this `dead' 
period, we compare $\sim$50 s of estimated background rate to the same duration of data, binned to 1.024 s resolution.  The 
goodness-of-fit can then be estimated by
\begin{equation}
	\chi^2 = \sum_{i=1}^N \frac{(C_i-B_i)^2}{C_i+\sigma_{B,i}^2},
\end{equation}
where $C_i$ is the number of observed counts in the $i$th bin and $B_i$ is the estimated background counts.  Dividing $\chi^2$ 
by the number of bins, $N$, gives the reduced chi-squared statistic.  Testing on data with known strong sources and occultation 
steps, a reasonable threshold to flag the background as problematic is $\chi^2/N=1.8$.  Very few false positive background errors 
are flagged with this threshold. Figure~\ref{BadBackground} shows an example where this method identifies a bias of the 
background estimation due to a nearby soft source.

The LB15 search is designed such that individual channels for individual detectors can be discarded if deemed necessary.  In the 
case of a poor background fit to one or two channels for a single detector, those channels would be removed from the likelihood 
calculation, but the rest of the data for that detector could be used.  Because both the null and alternate hypotheses in the 
likelihood ratio calculation use the same data for a time of interest, flagging particular channels for removal from the analysis is 
statistically valid.  The LB15 search, however, employs a likelihood that looks for coherent and consistent signal over the entire 
GBM energy band in all detectors, therefore if too many energy channels from a detector or multiple detectors are removed, the 
likelihood of a signal over background will be significantly diminished.

\subsection{Removing Sharp SAA Rate Changes}
\begin{figure}
	\begin{center}
		\includegraphics[scale=0.8]{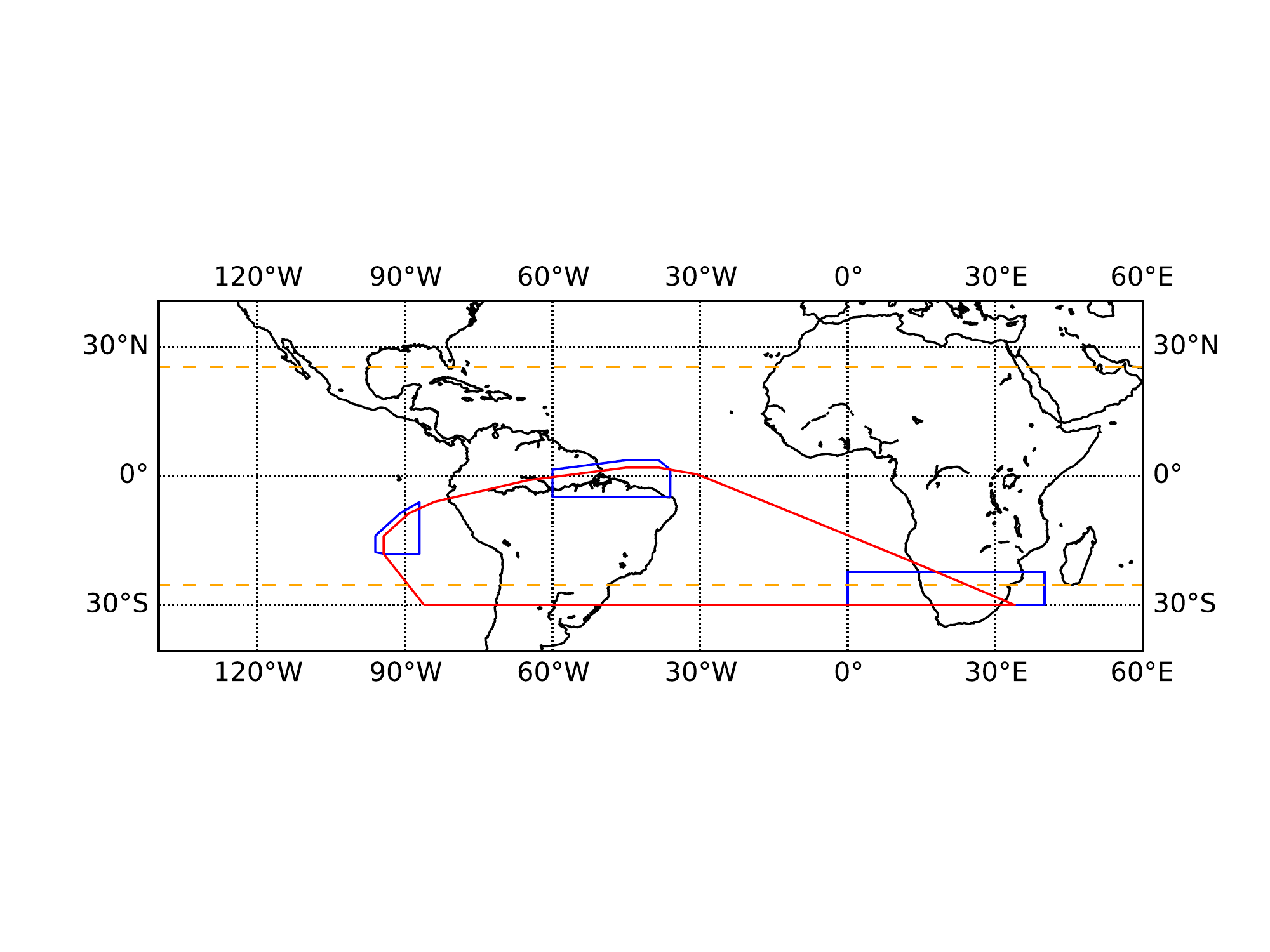}
	\end{center}
\caption{The SAA region where the GBM detectors are disabled is marked in red. The extended SAA regions that we use for event 
filtering are marked by the blue polygons.  The yellow dashed lines show the North-South extent of the {\it Fermi} orbit.
\label{SAA}}
\end{figure}

Although the background estimation detailed in Section 2 can model the background during approach and exit from the SAA 
(albeit with larger uncertainty), there may be particular times when the rate increase/decrease is too steep during periods of high 
particle activity when the true region of the SAA expands or geographically shifts beyond the SAA polygon stored in the {\it Fermi} 
flight software.  For this reason, it is desirable to implement an extended SAA polygon and filter out any events that happen 
when {\it Fermi} is within the expanded polygon, which is shown in Figure~\ref{SAA}.  This filter can be implemented outside of the 
LB15 search so that if a time of interest is within the expanded polygon (but the GBM detectors are still collecting data), a search 
can still be performed without the filter at the cost of a higher FAR.

\subsection{Likelihood Calculation Pre-Filter\label{sec:Prefilter}}
The likelihood ratio calculation, as described in LB15, determines the likelihood that a candidate is a real signal as opposed to a 
background fluctuation.  This calculation requires an estimation of the signal amplitude that maximizes the likelihood of the 
alternate hypothesis that a signal exists (Eq. 13 of LB15).  The signal amplitude solution must be solved numerically as there 
exists no closed-form analytical solution.  The LB15 search performed this numerical optimization using Newton's method on every 
single time bin searched.  This evaluation is the costliest part of the LB15 search, so we have included a pre-filter to the numerical 
calculation so that initial guess log-likelihood ratio values of $\le5$ are not treated to the full numerical optimization.  This value is 
chosen because it is well below the significance threshold for a candidate signal.  We have tested to ensure that the initial guess 
log-likelihood ratios $\le5$ do not become `significant' candidate detections.  Implementing this pre-filter has produced up to a 
factor of $\sim5$ increase in speed.

%% Joint Spatio-Temporal Ranking Statistic
\section{Joint Spatio-Temporal Ranking Statistic\label{sec:Statistic}}
The LB15 search during O1 used the LogLR as the ranking statistic for significance, specifically Equation 20 in LB15, which a
assumes a uniform sky prior.  We now choose to include a location prior in the form of a LIGO sky map. For this we use 
equation 21 from LB15:
\begin{equation}
\Lambda_{GW-GBM} = \int{d\Omega P_{GW}(\Omega) \Lambda_{GBM}(\Omega)},
\end{equation}
where $P_{GW}(\Omega)$ is the LIGO localization used as a prior probability distribution on the sky and $\Lambda_{GBM}$ is the 
GBM likelihood ratio which is also a function of sky position, among other dependencies.  
%We are currently exploring adding a 
%threshold based on the difference between this modified likelihood ratio, CoincLR, and the LogLR assuming a uniform sky prior.

%%Trivial Code Modifications
\section{Trivial Code Modifications}
The code was updated from Python 2 to Python 3, as well as updating the modules and functions to modern versions. An incorrect 
sign in the reference log-likelihood was corrected (this has no effect on any of the FAR calculations as all log-likelihood ratios were 
modified by the same constant value). The original version of the code was written to run on LIGO clusters some time ago and, as 
a result, optimized memory usage. We have removed most of this in favor of speed.

A parallelized version of the polynomial interpolation that numerically calculates the log-likelihood value was implemented using 
the {\bf Theano} module~\citep{Theano16}. This implementation, along with the switch to Python 3, produces identical results to
the LB15 search.

%%Validation
\section{Validation}
We have validated the implemented changes by testing the search on triggered short GRBs, simulated GRBs injected into real 
GBM data, and on background with no known sources to estimate the change in the FAR.

%%Validation with Signals
\subsection{Validation with Signals\label{sec:SignalValidation}}
\begin{figure}
	\begin{center}
		\includegraphics[scale=0.8]{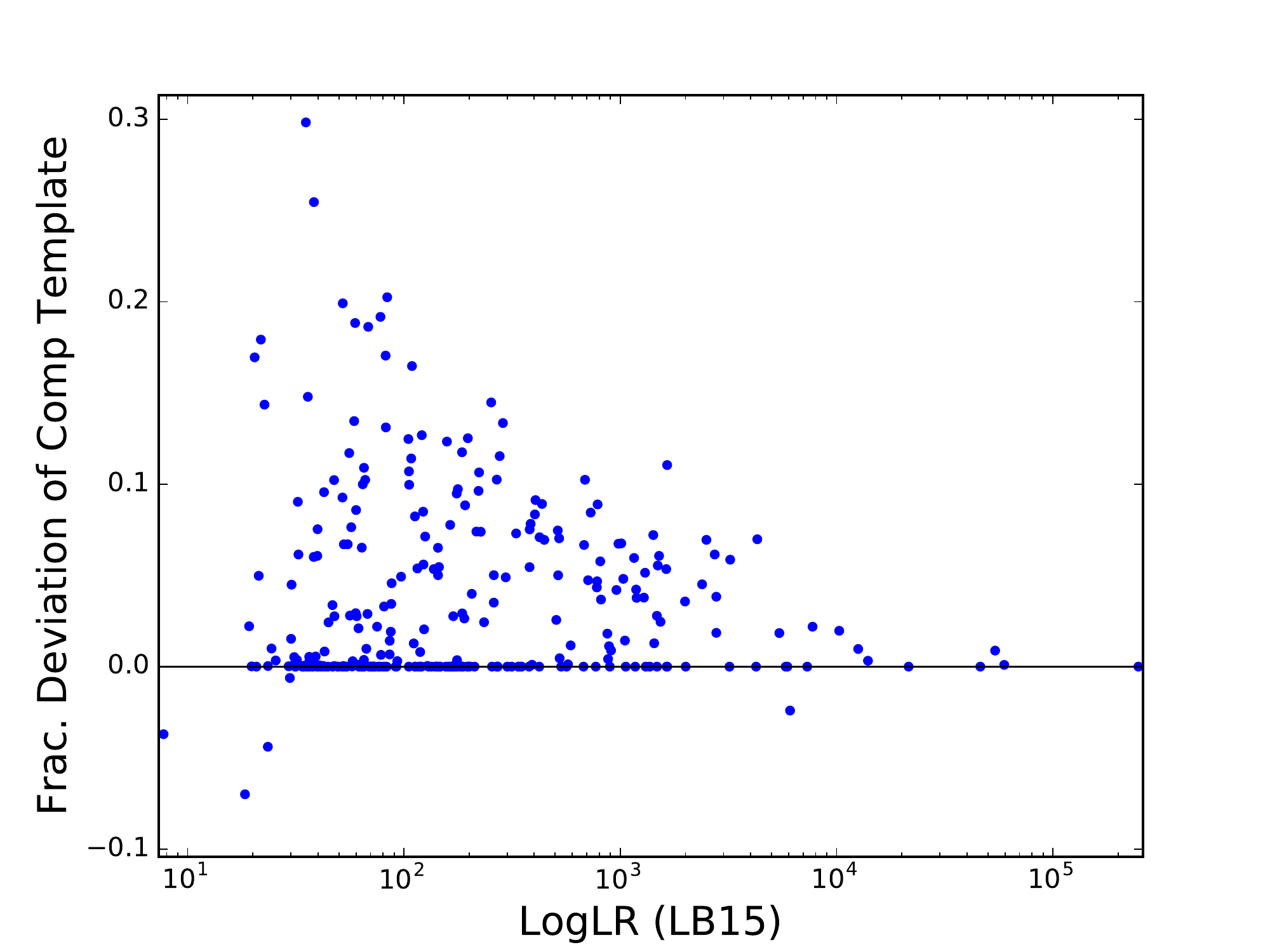}
	\end{center}
\caption{Comparison of the detection statistic when using the original hard Band function template compared to the new 
Comptonized function template for 310 GBM-triggered short GRBs.  The GRBs that lie on the unity line are mostly found with the 
`Normal' template instead of the hard or Comptonized templates.
\label{AllSGRBs}}
\end{figure}

Using 310 short GRBs that triggered GBM, we compared the LB15 using the hard spectral template to the replacement template 
using the hard Comptonized function.  The new template increased the ranking statistic in 84\% of the cases, and was equivalent 
for 14\% of the cases, with an overall average 4\% increase in the statistic value compared to the original.  The increase in the 
statistic is more significant for lower values of the statistic, where we expect un-triggered short GRBs to be uncovered using this 
search.  The comparison is shown in Figure~\ref{AllSGRBs}.

\begin{figure}
	\begin{center}
		\subfigure[]{\label{LocSGRBsOffset}\includegraphics[scale=0.5]{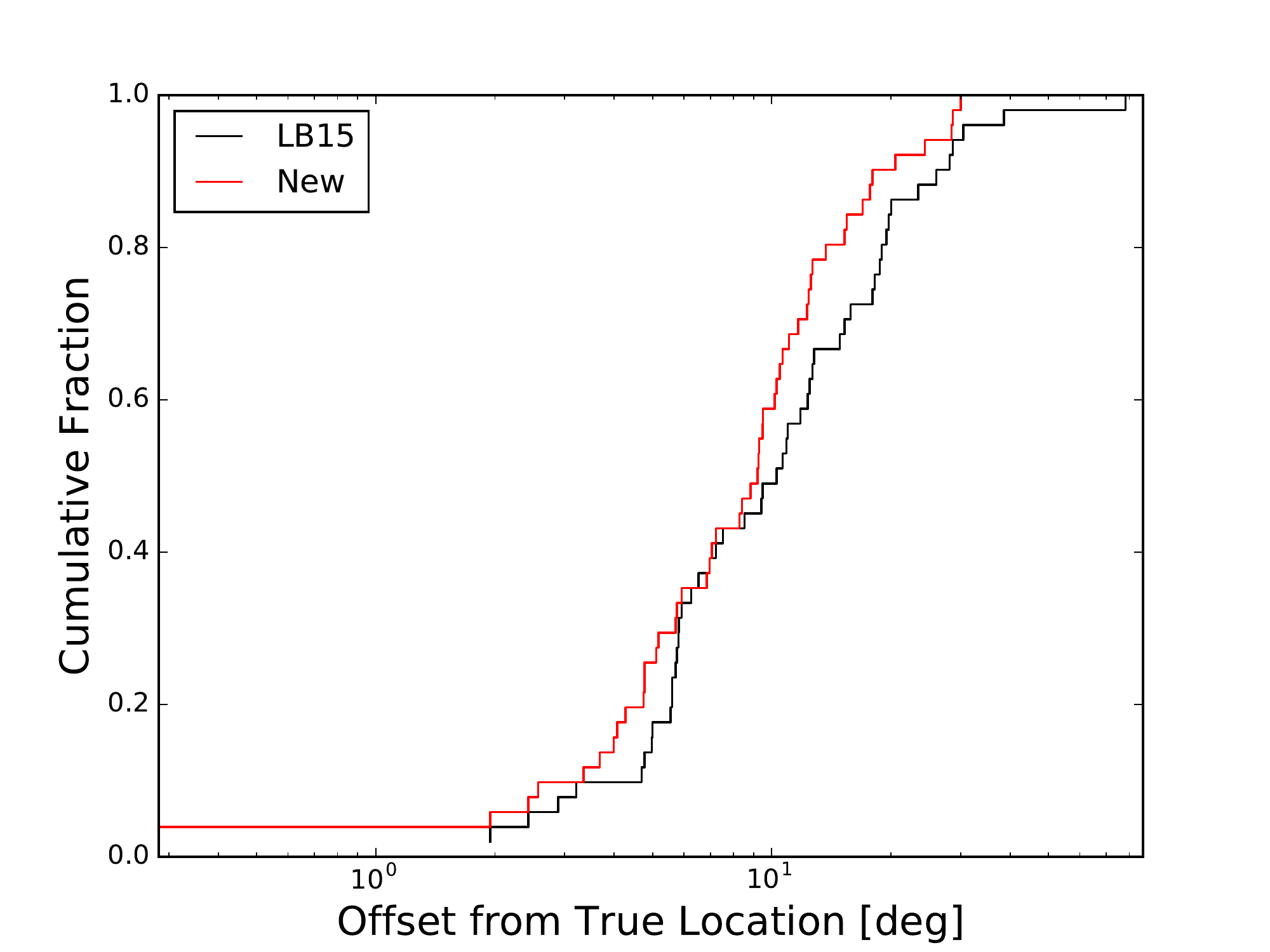}}\\
		\subfigure[]{\label{LocSGRBsCoincLR}\includegraphics[scale=0.5]{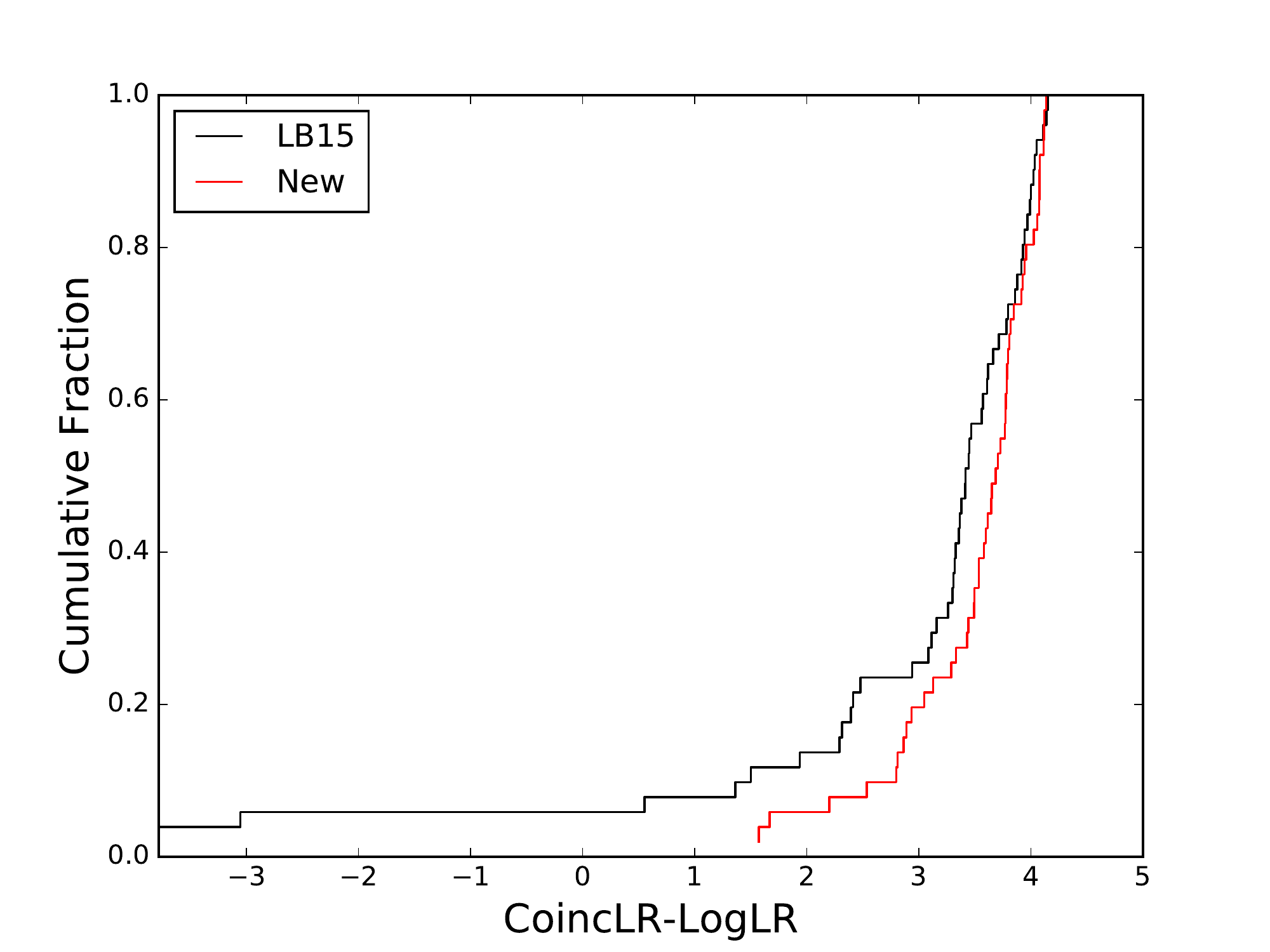}}
	\end{center}
\caption{Comparison of the~\ref{LocSGRBsOffset} offset of the maximum likelihood location from the true location and 
the~\ref{LocSGRBsCoincLR} coincidence ranking statistic between the LB15 search templates and the new Comptonized 
template for 51 triggered short GRBs with known location.
\label{LocSGRBs}}
\end{figure}

Additionally, we ran the search on 51 triggered short GRBs with known sky location (i.e. localized by {\it Swift} or the {\it Fermi} 
LAT).  The search was run using a Gaussian prior on the sky centered at the true location with a 1$\sigma$ radius of 10 degrees.  
Ideally if the search correctly localizes a truly associated candidate event with a specific sky prior, then the coincident ranking 
statistic described in Section~\ref{sec:Statistic} should be larger than the ranking statistic resulting from a uniform sky prior.  In 
Figure~\ref{LocSGRBs} we show the comparisons between the LB15 search and the search using the new Comptonized template 
for both the offset of the maximum likelihood location from the true location and the coincident ranking statistic.  The offset from 
the true location is improved for 59\% of the short GRBs that are found by the hard or Comptonized templates, while the increase 
in the coincident ranking statistic is more significant for 68\% of the GRBs when using the Comptonized template.

\begin{figure}
	\begin{center}
		\subfigure[]{\label{LocInjOffset}\includegraphics[scale=0.5]{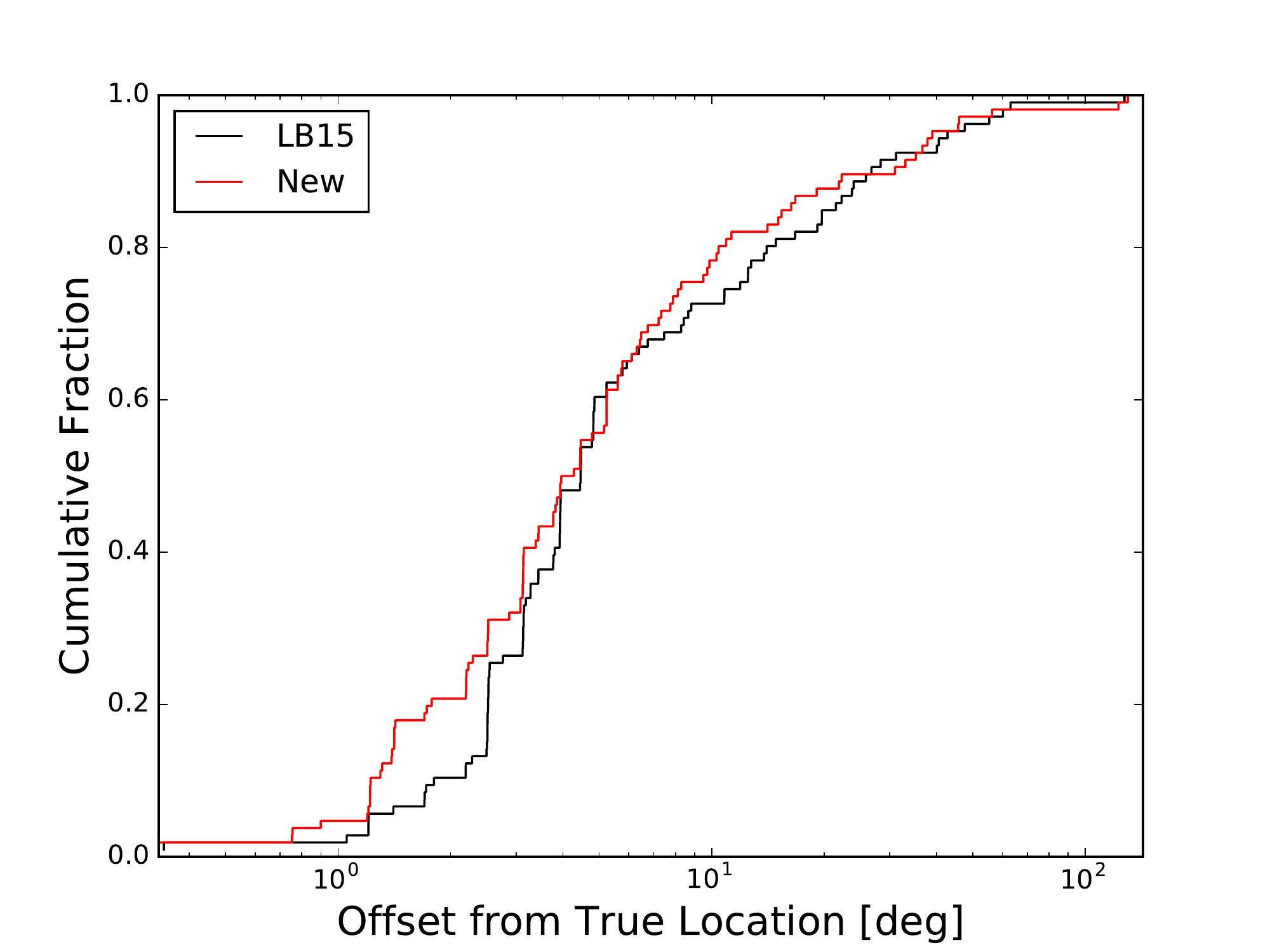}}\\
		\subfigure[]{\label{LocInjCoincLR}\includegraphics[scale=0.5]{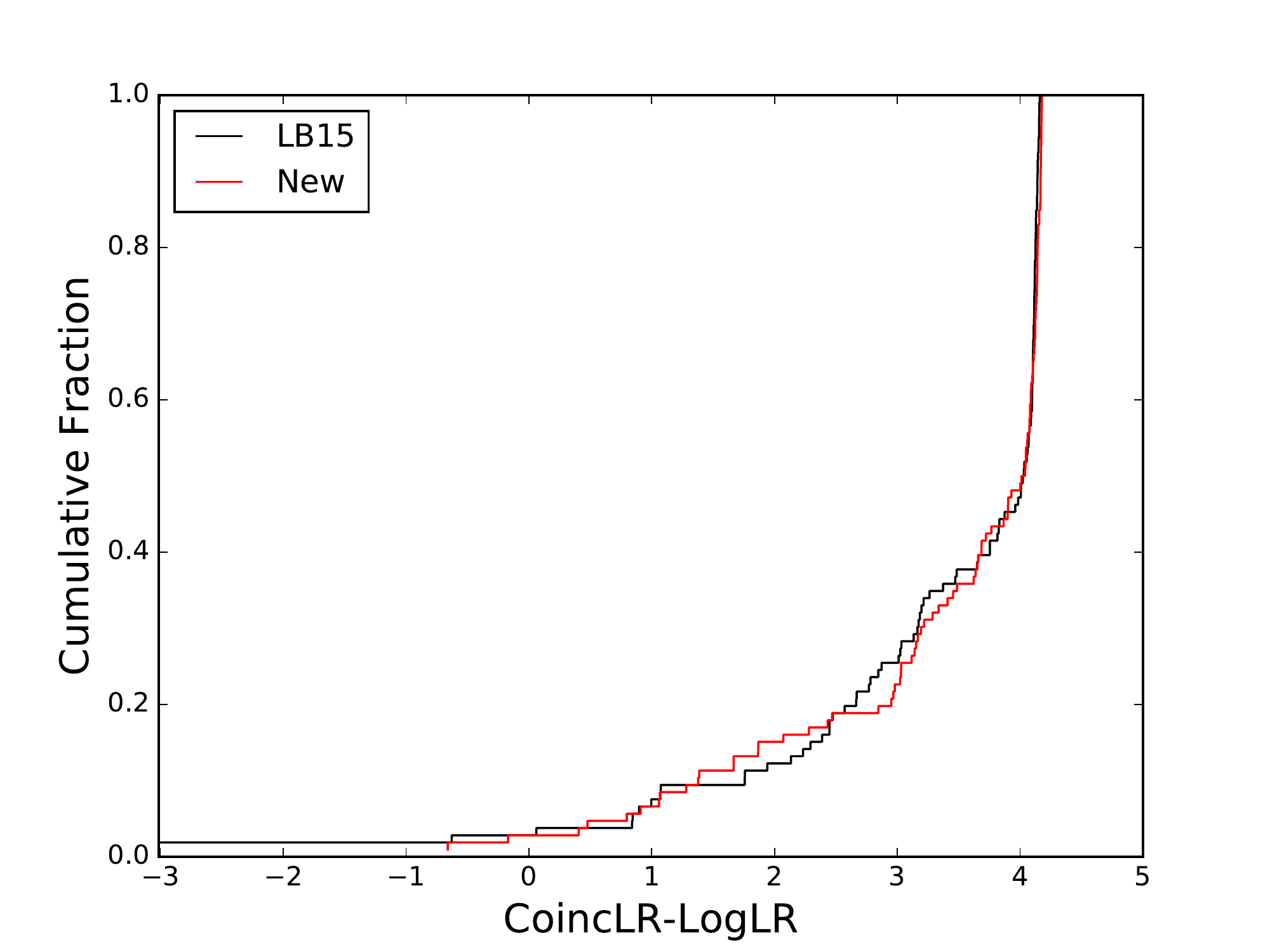}}
	\end{center}
\caption{Comparison of the~\ref{LocInjOffset} offset of the maximum likelihood location from the true injected location and
the~\ref{LocInjCoincLR} coincidence ranking statistic between the LB15 search and the updated search for 105 injected GRBs 
found by both searches.
\label{LocInjections}}
\end{figure}

Another source test that we have performed is simulating weak short GRBs with a variety of durations, spectra, and locations 
around the spacecraft and injecting them into real GBM background.  By doing this, we can see how often we can detect the 
injected GRB and accurately localize them.  A full estimation of the recovery efficiency as a function of spectrum, location, and 
duration is ongoing, but we show a sample of tests to compare the LB15 search to the implemented changes.  
Figure~\ref{LocInjections} shows the comparison of the true offset from the injected location and the coincidence ranking statistic 
between the LB15 search and the implemented template and background improvements.  In this particular test, 200 GRBs were 
simulated with varying spectra and locations, and at varying brightness close to the detection threshold.  Both searches 
reasonably recovered 112 of 200 injections, and Figure~\ref{LocInjections} shows the comparison of the 105 injections that were 
recovered by both searches.  The upgraded search recovered a maximum likelihood location closer to the true injected location 
56\% of the time compared to a smaller offset 33\% of the time for the LB15 search.  The increase in the spatial coincidence 
ranking statistic was greater with the implemented upgrades for 60\% of the recovered events and the other 40\% were better with 
the original LB15 search.

\begin{figure}
	\begin{center}
		\subfigure[]{\label{CumulCoincLR}\includegraphics[scale=0.5]{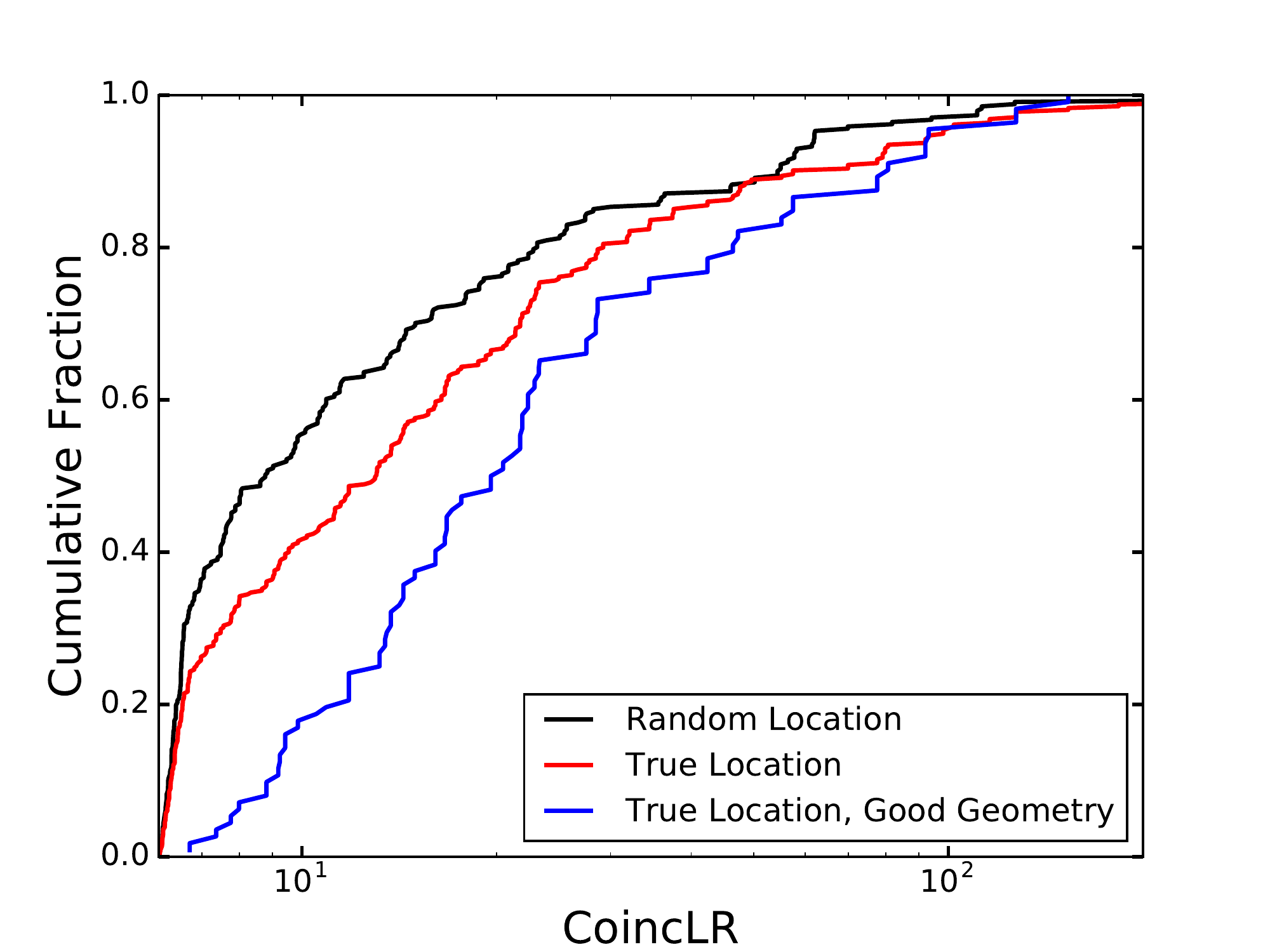}}\\
		\subfigure[]{\label{FDevCoincLR}\includegraphics[scale=0.5]{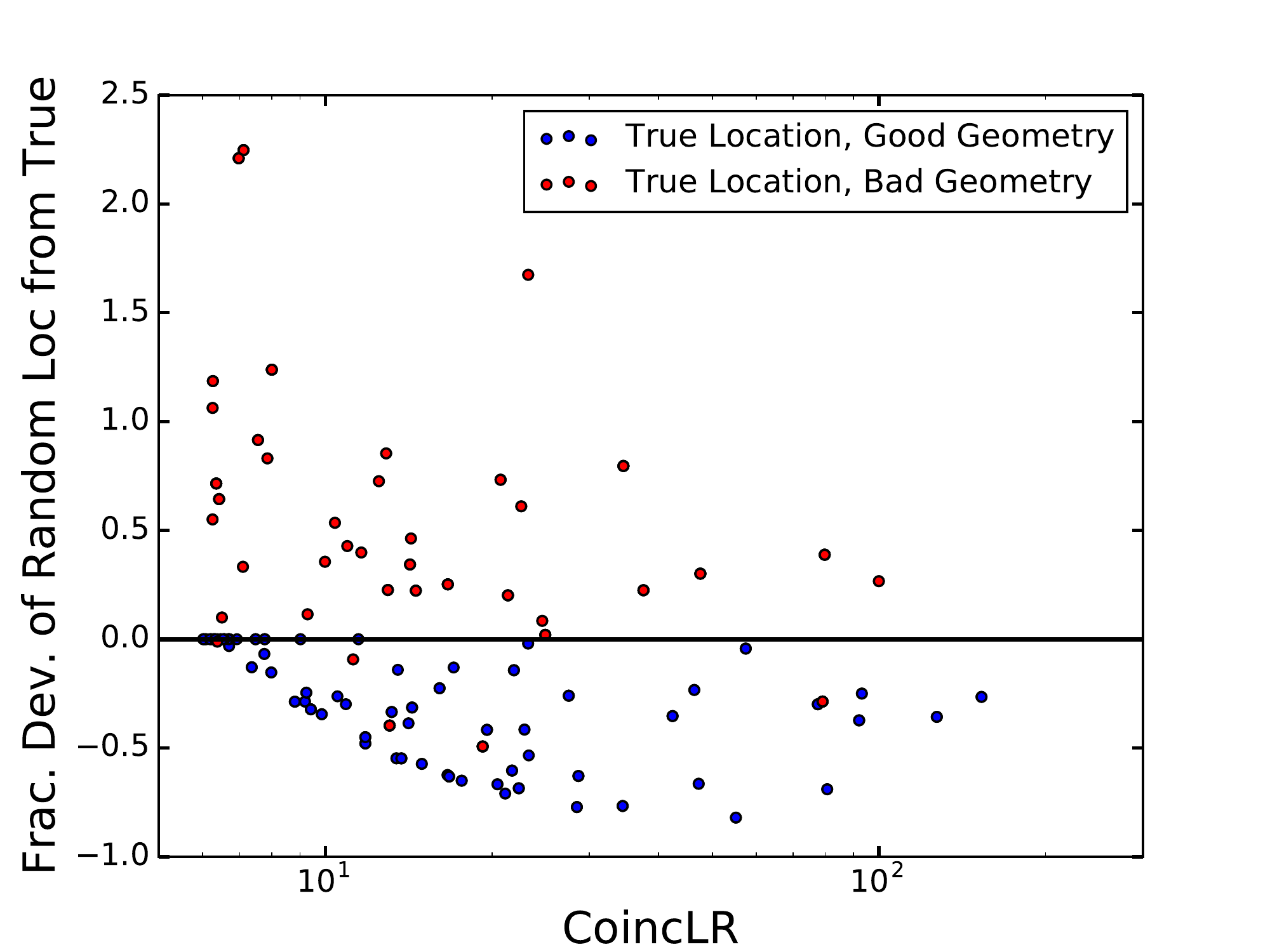}}
	\end{center}
\caption{Panel~\ref{CumulCoincLR} shows the cumulative distributions of the CoincLR ranking statistic for the injected GRB 
signals at the LIGO sky map locations (red) and injected signals at random locations on the sky (black).  The blue distribution 
shows the CoincLR for the injected GRBs that have a more favorable geometry to the spacecraft compared to the randomly 
injected sky locations.  Panel~\ref{FDevCoincLR} shows the fractional deviation from the injected CoincLR for a random injection 
on the sky. \label{BayesInjections}}
\end{figure}

We performed a final test using a set of simulated weak short GRBs injected at times and sky locations consistent with simulated 
LIGO BAYESTAR localization maps~\citep{Singer14}.  The times at which the BAYESTAR maps were simulated were shifted in 
time such that the LIGO detectors were in the same geometry relative to the simulated originating sky position and during times 
that GBM had continuous TTE data.  We also created a test sample of the simulated short GRBs injected at the same times 
and with precisely the same spectral shape, but with randomized locations on the sky.  We then used the sky maps as priors in the 
search to determine the effectiveness of providing spatial information to the search in the event that a real signal is detected.  
Figure~\ref{BayesInjections} shows the comparison of the CoincLR ranking statistic for the two injection sets.  The injections 
consistent with the BAYESTAR maps generally have a larger CoincLR than the random injections, however since the strength of a 
signal in the GBM data results from a convolution of the photon model with the highly-angular-dependent detector responses, 
there are cases where the randomly injected location produces a stronger signal in the data than a signal that is consistent with 
the sky prior.  In these cases, the LogLR, the ranking statistic resulting from a uniform sky prior, is also larger than that for the 
BAYESTAR-consistent injections.  Figure~\ref{BayesInjections} demarcates, in blue, the distribution of injected events for which 
the BAYESTAR-consistent injections have at least comparable signal strength to the randomly injected signals based 
on LogLR, which is $\sim$67\% of the injected signals.  All randomized sky injections that have a higher ranking statistic compared 
to the BAYESTAR-consistent injections are observed at comparatively better geometries to the detectors.

\begin{figure}
	\begin{center}
		\subfigure[]{\label{ShiftOne64}\includegraphics[scale=0.5]{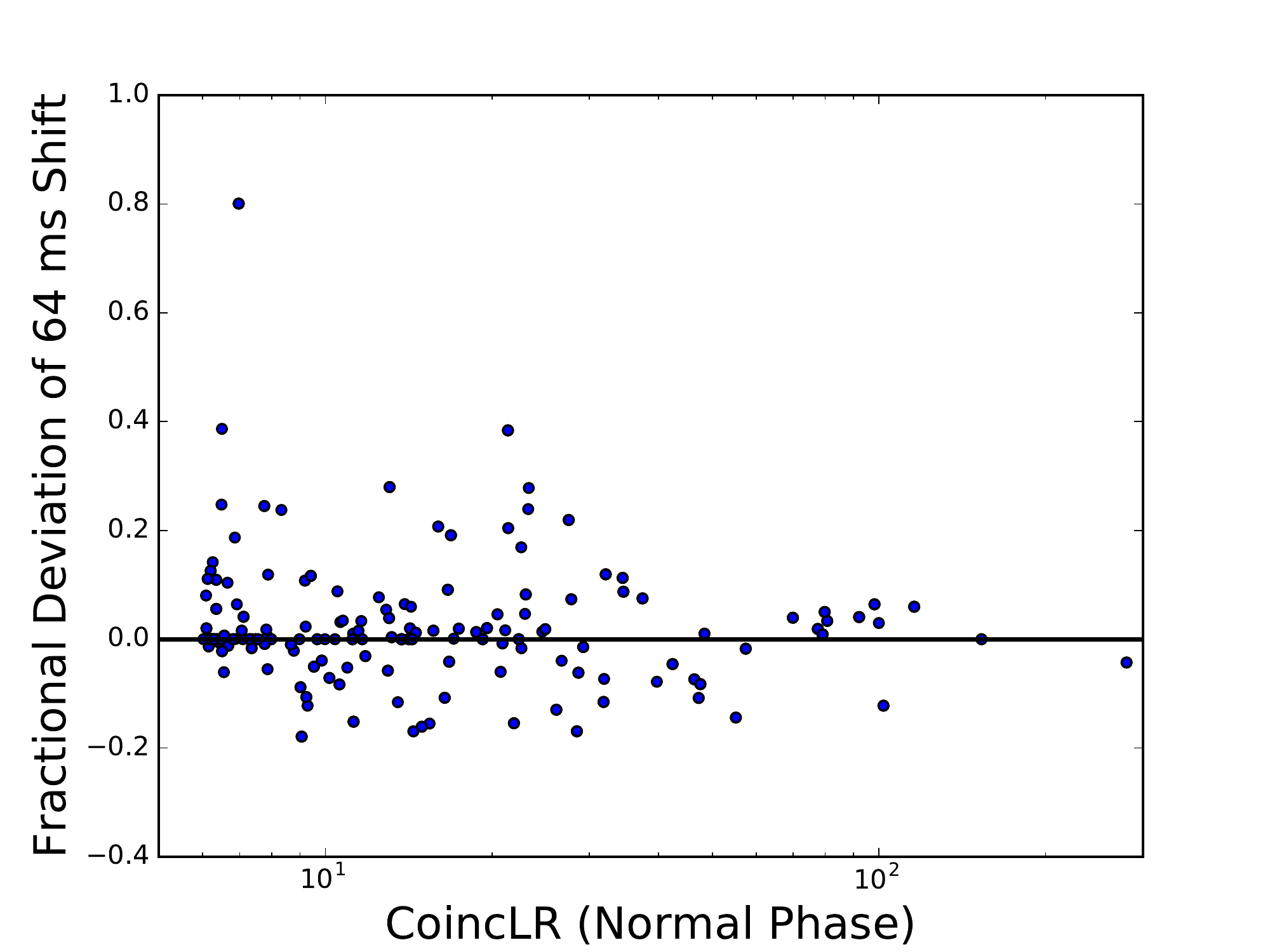}}\\
		\subfigure[]{\label{ShiftEach64}\includegraphics[scale=0.5]{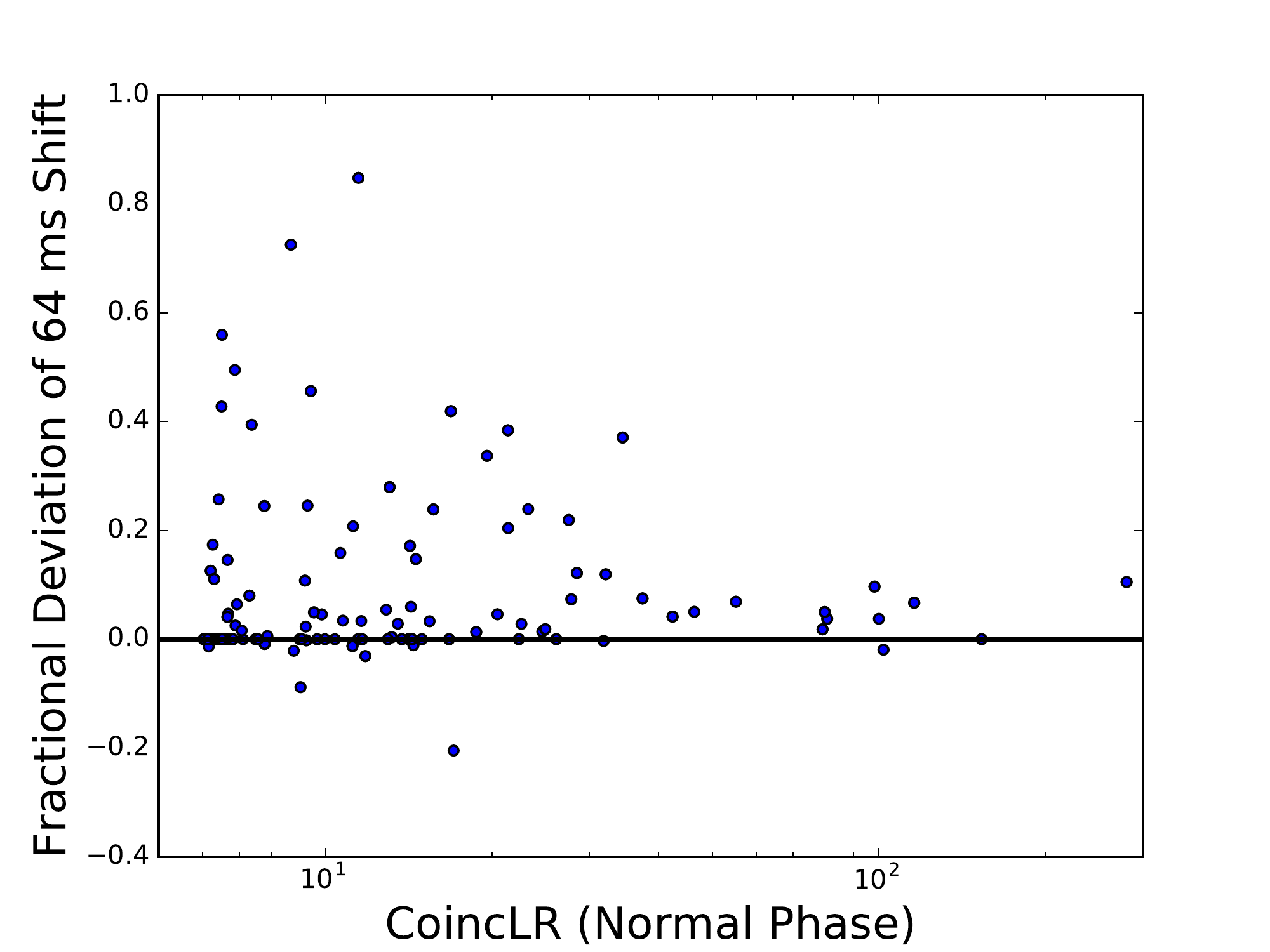}}
	\end{center}
\caption{Panel~\ref{ShiftOne64} shows the CoincLR ranking statistic of injected signals with the search window centered on the 
time of interest (Normal Phase), compared to a search window shifted by 64 ms from the time of interest. Panel~\ref{ShiftEach64} 
shows the Normal Phase CoincLR compared to the CoinCLR when utilizing a phase shift factor of 16 for all timescales.\label{Shift64}}
\end{figure}

As a final note on the validation of the updated search using injected signals, we investigated the impact of performing the search 
with a maximum factor 4 phase shift relative to the search timescale as discussed in Section~\ref{sec:InputData}, rather than a 
constant 64 ms phase shift.  The targeted search is based on searching a time window centered on a particular time of interest 
(what we refer to here as Normal Phase), so we re-ran the search on the BAYESTAR-consistent injections but shifted the center of 
the time window by 64 ms relative to the time of interest.  The comparison of the ranking statistic resulting from this shift is shown 
in Figure~\ref{ShiftOne64}.  Because search timescales $>$ 256 ms are not utilizing the ability to perform a 64 ms phase shift of 
the binned data, the resulting CoincLR can change by a considerable amount (an average of $\sim$10\% here), which can result 
in a large change in the associated FAR for a particular event. This implies that the current search, as devised for LIGO O2a, 
requires some amount of ``luck" to find a weak, short signal because of this phasing issue.  If, on the other hand, we run the 
search allowing for a maximum phase shift factor of 16 for all timescales, the case improves considerably as shown in 
Figure~\ref{ShiftEach64}, where the CoincLR increases in nearly every case, and the overall average change is a $\sim$9\% 
increase in CoincLR over the search using the Normal Phase.  Increasing the number of phase shifts for the longer timescales 
effectively increases the size of the search space, therefore the FAR would have to be recalculated.  Because the FAR estimation 
is very time-consuming (and even more so if the factor 16 phase shift for all timescales were implemented) we leave this potential 
change for future study and continue with the factor 4 phase shift for LIGO O2a.

%%Validation with Background
\subsection{Validation with Background}
\begin{figure}
	\begin{center}
		\subfigure[`Normal' Template]{\label{NewBackMedium}\includegraphics[scale=0.31]{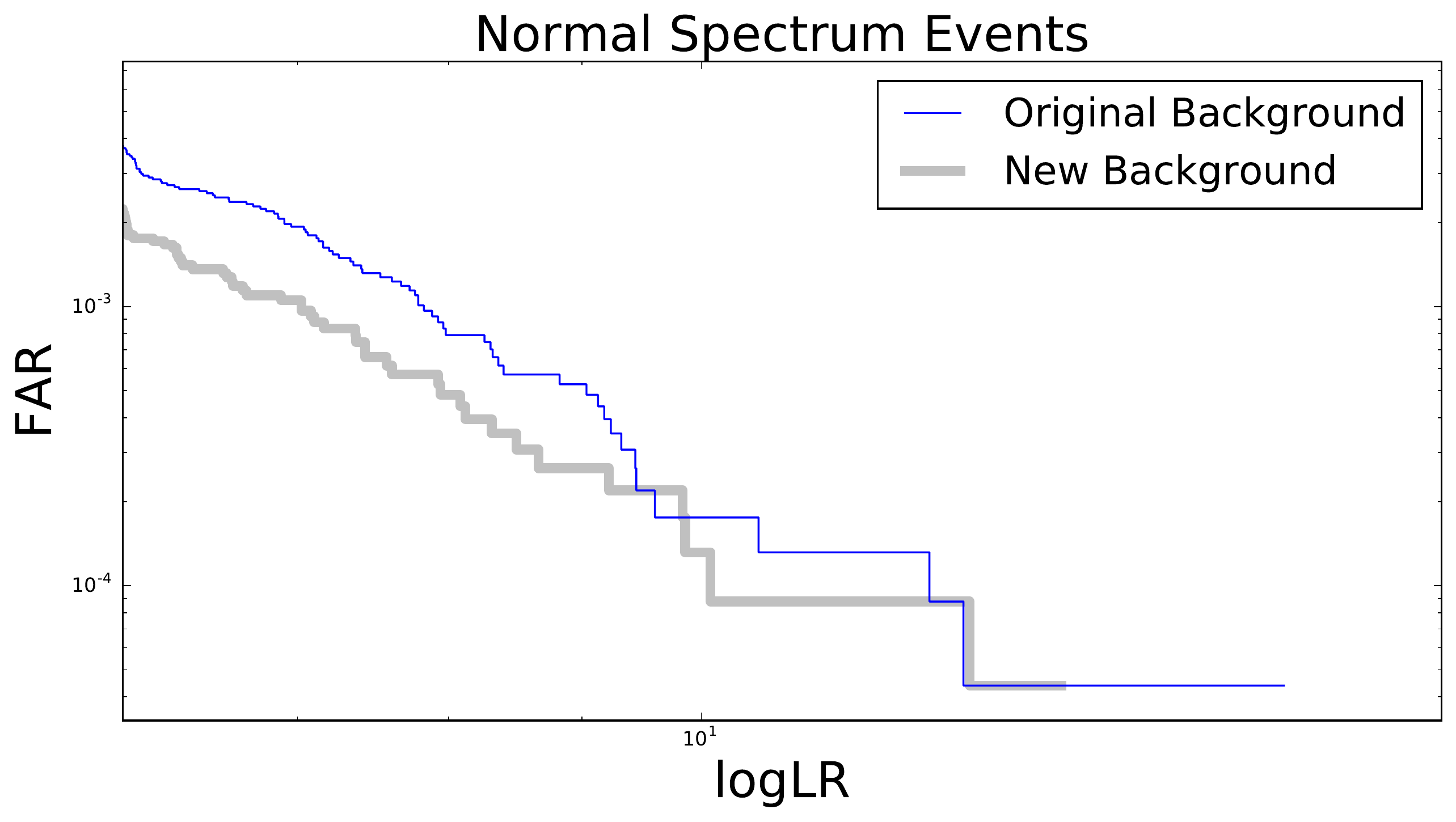}}
		\subfigure[`Hard' Template]{\label{NewBackHard}\includegraphics[scale=0.31]{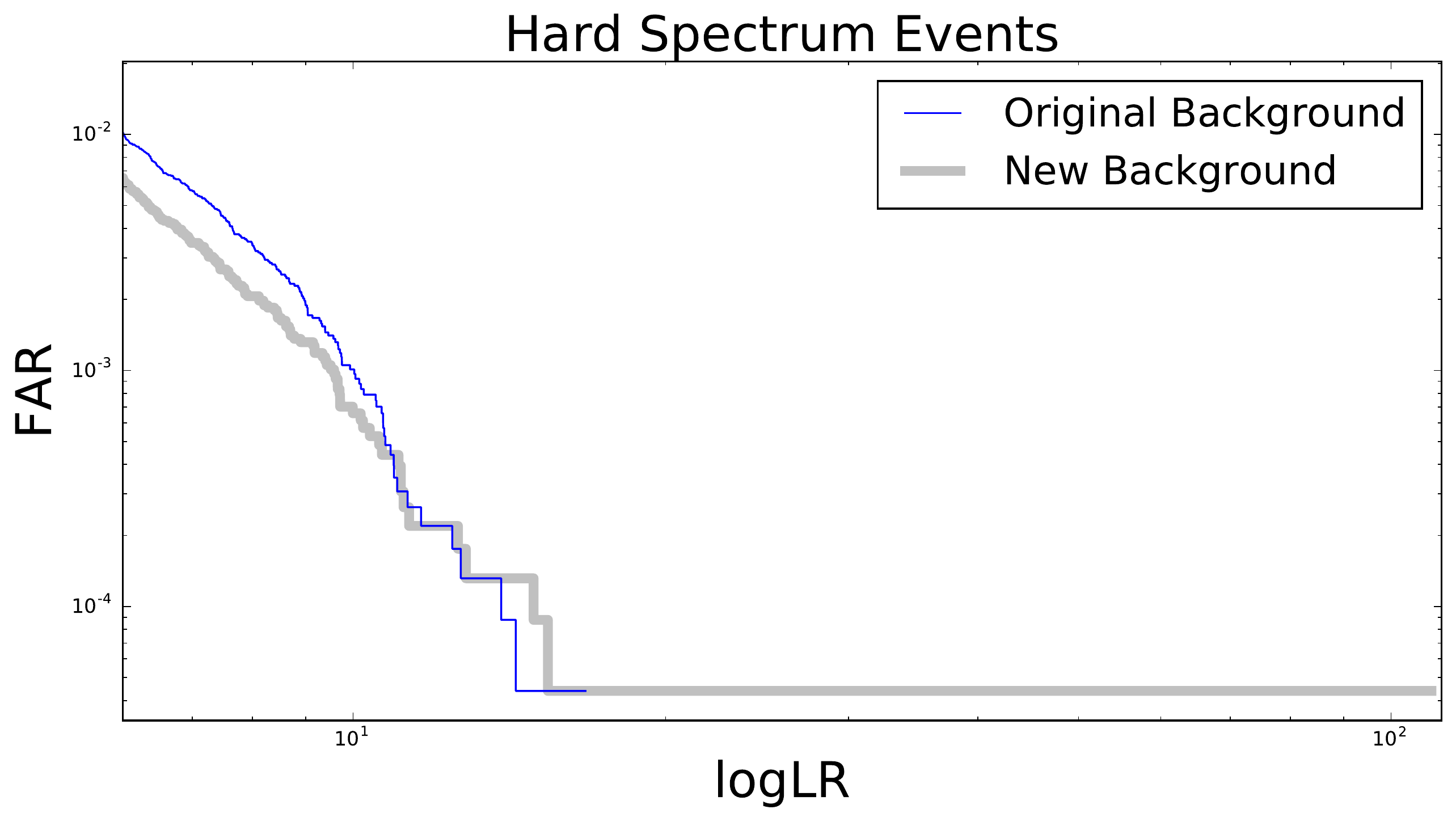}}\\
		\subfigure[`Soft' Template]{\label{NewBackSoft}\includegraphics[scale=0.31]{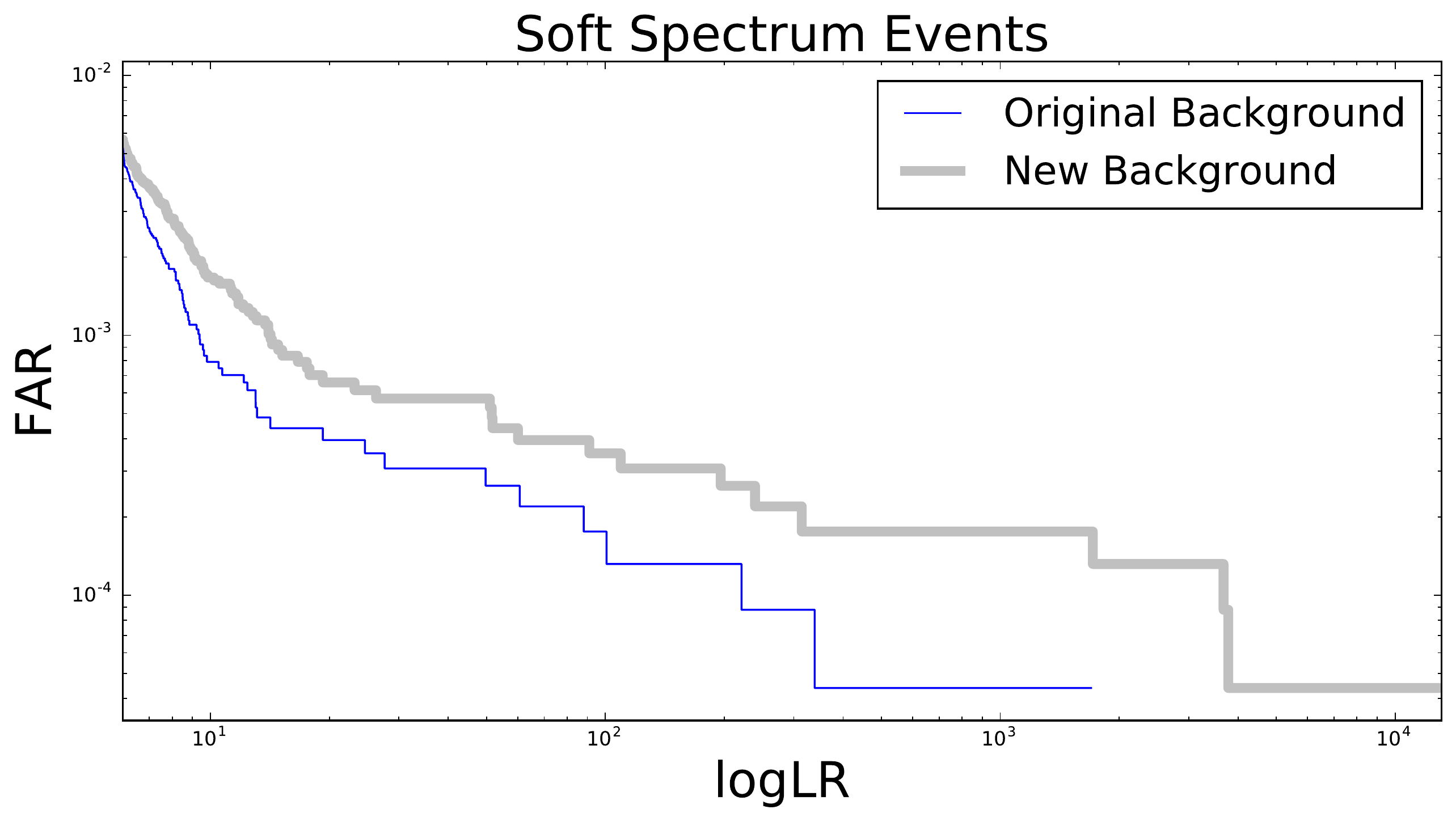}}
		\subfigure[`Soft' Template $> 2$ s]{\label{NewBackSoft2}\includegraphics[scale=0.325]
		{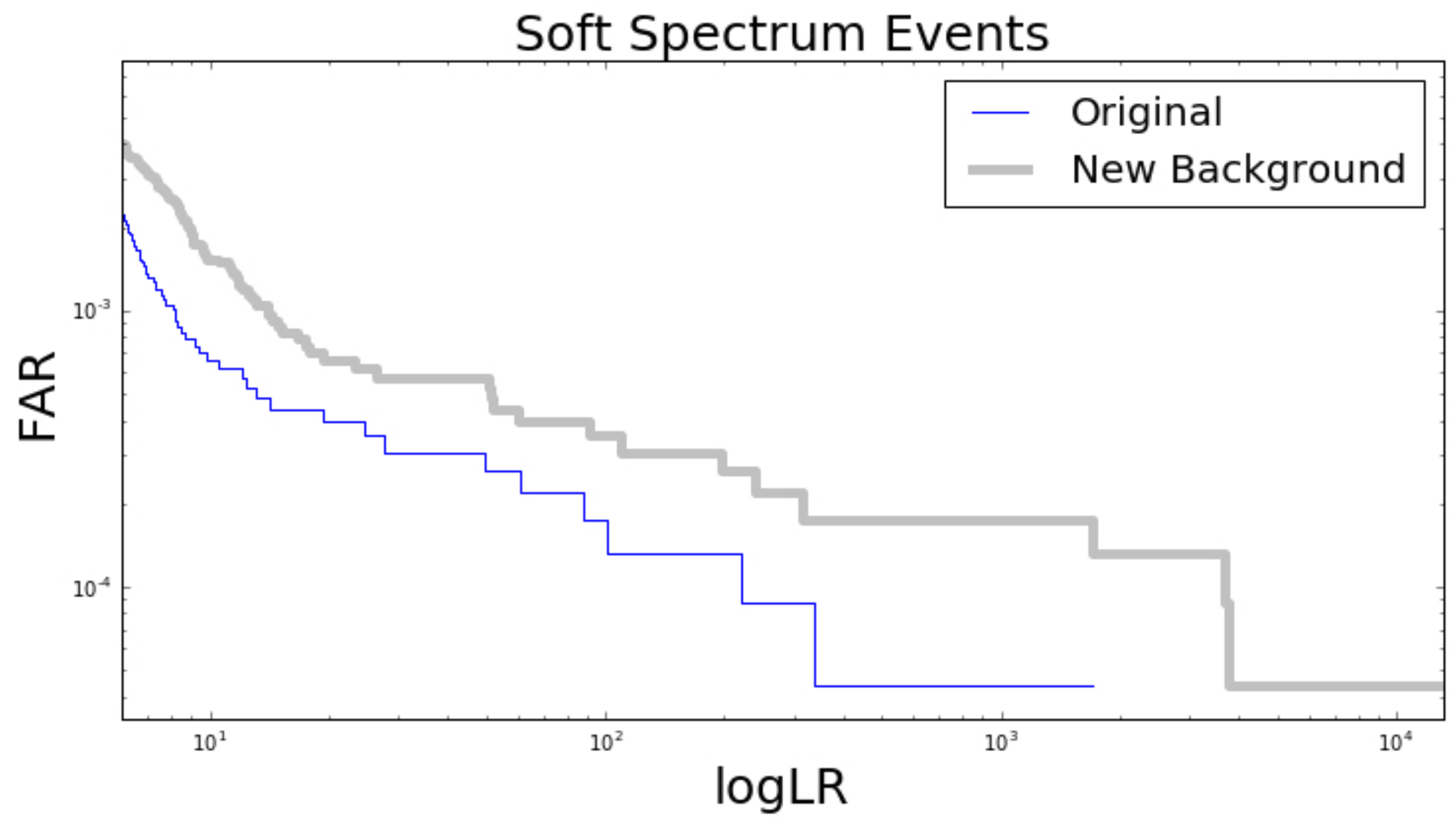}}
	\end{center}
\caption{The FAR comparison between the LB15 search and the new background estimation described in 
Section~\ref{sec:NewBack} for each of the three original spectral templates used in LB15.  Panels~\ref{NewBackMedium}--
\ref{NewBackSoft} show the FAR combining all candidate timescales from 0.256~s up to 8.192~s.  Panel~\ref{NewBackSoft2} 
shows the FAR for the soft template only at timescales $> 2$ s, indicating that the overrun in the new FAR compared to LB15 is 
due to longer timescale candidates.
\label{NewbackFAR}}
\end{figure}

\begin{figure}
	\begin{center}
		\includegraphics[scale=0.8]{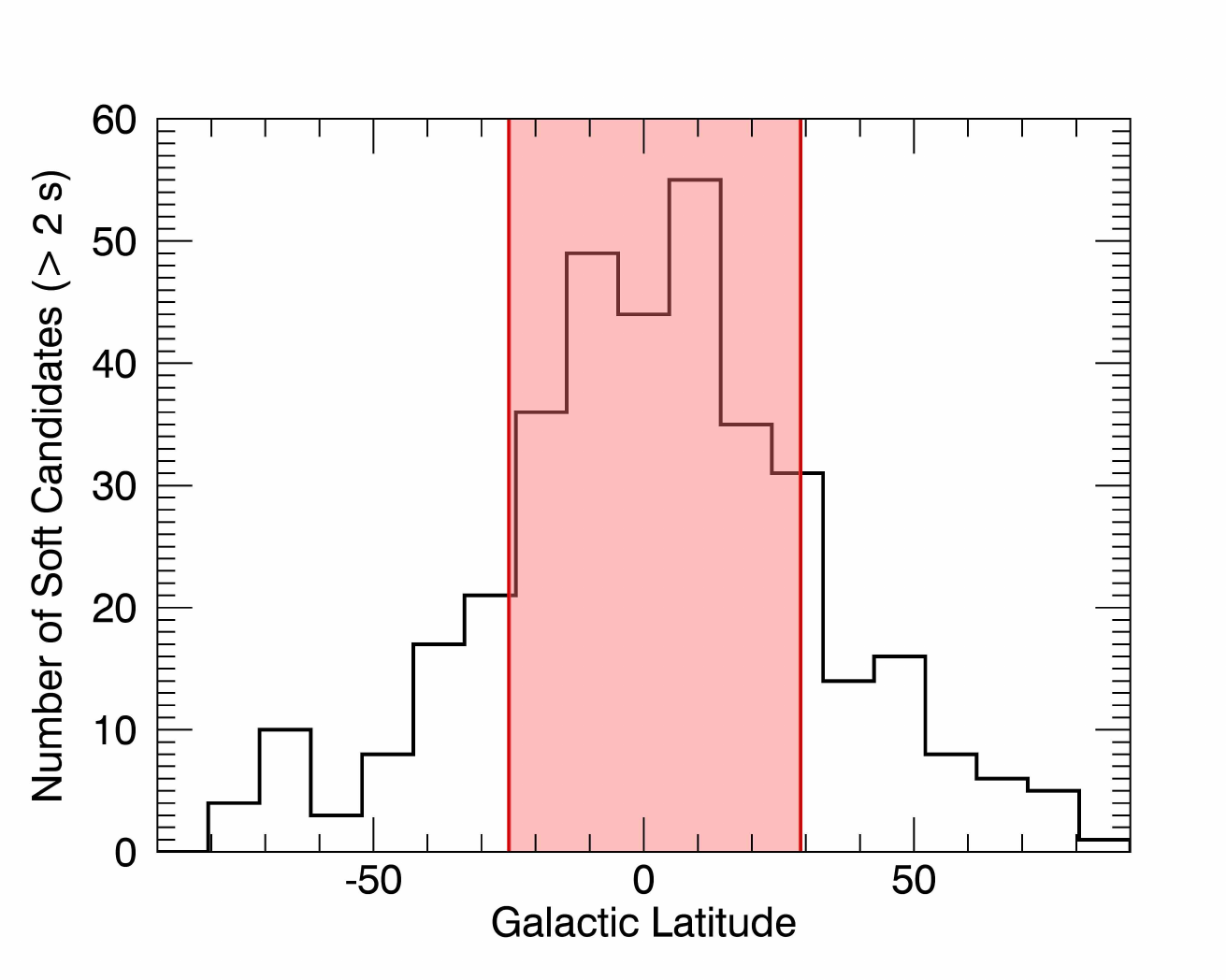}
	\end{center}
\caption{The Galactic latitude distribution of 363 significant soft candidates at $>2$~s timescale using the new 
background estimation.  The red shaded region represents the angular extent of 67\% of the candidates.  Note that only the 
maximum likelihood location was considered, and this distribution does not take into account localization uncertainty.
\label{NewBackSoftMap}}
\end{figure}

In addition to testing the changes with real GRBs and injected signals, we test over regions of background where candidate 
signals are not expected (around the times of the BAYESTAR injections in the previous section) to evaluate the new FAR and 
compare it to the FAR of the LB15 search.  Figure~\ref{NewbackFAR} shows the FAR comparison using the new background 
estimation for each of the three templates.  For both the `Normal' and the `Hard' templates, the FAR with the new background 
clearly under-runs the FAR from LB15, thereby making the detection of candidates more significant.  The FAR for the `Soft' 
template clearly over-runs compared to LB15, but as can be seen in Figure~\ref{NewBackSoft2}, the cause of the over-run is due 
to events that are $> 2$ s in duration.  One possible explanation for this over-run is a more efficient recovery of spectrally soft 
Galactic sources that may be flaring in the low-energy channels of the GBM data.  Indeed, if we map the maximum likelihood 
localizations of all the events in Figure~\ref{NewBackSoft2}, we find that a large fraction of events are consistent with a Galactic 
origin, shown in Figure~\ref{NewBackSoftMap}.  This implies that the FAR for the `Soft' template is bounded by the amount of soft 
Galactic activity GBM observes, and we are currently investigating using a Galactic sky prior to classify soft events that strongly 
localize to the Galactic Plane as well as dividing the FAR distributions into different timescales.  Another class of events that can 
contribute to the longer duration soft events are solar flares, which are usually easy to distinguish from astrophysical events by 
inspection of the GBM lightcurves and contemporaneous GOES data.   

\begin{figure}
	\begin{center}
		\subfigure[`Normal' Template]{\label{NewTempMedium}\includegraphics[scale=0.31]{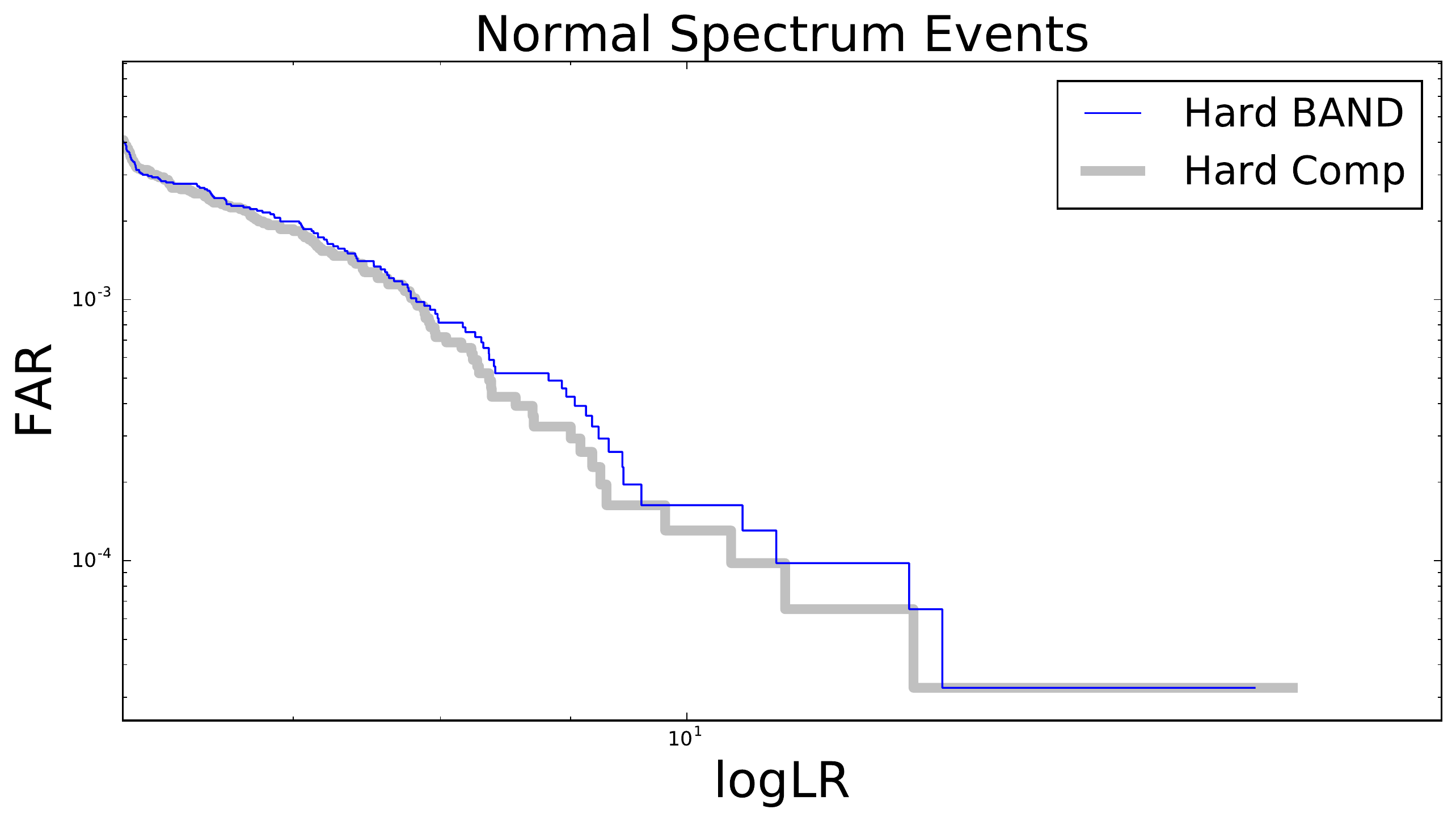}}
		\subfigure[`Hard' Template]{\label{NewTempHard}\includegraphics[scale=0.31]{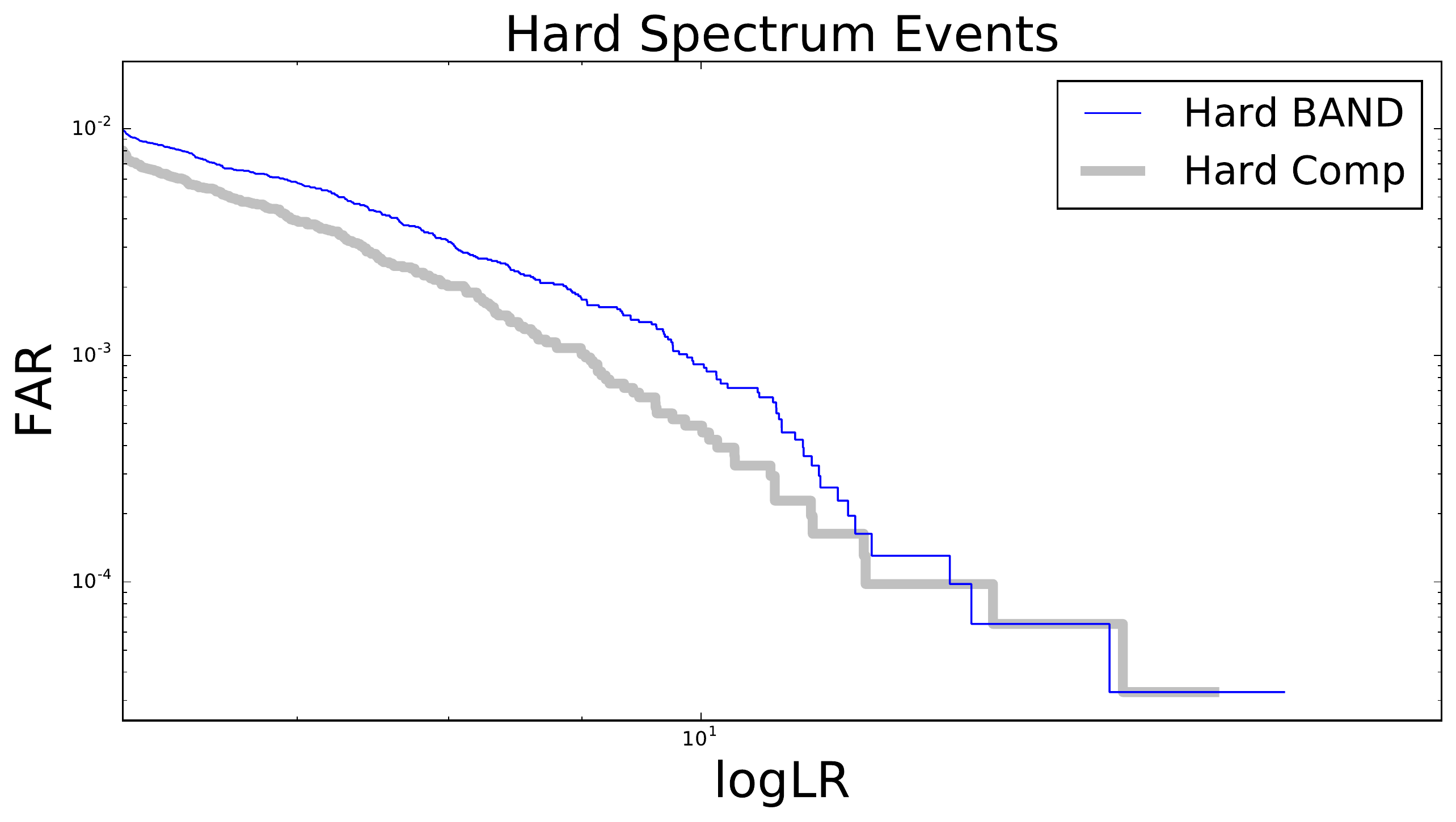}}\\
		\subfigure[`Soft' Template]{\label{NewTempSoft}\includegraphics[scale=0.31]{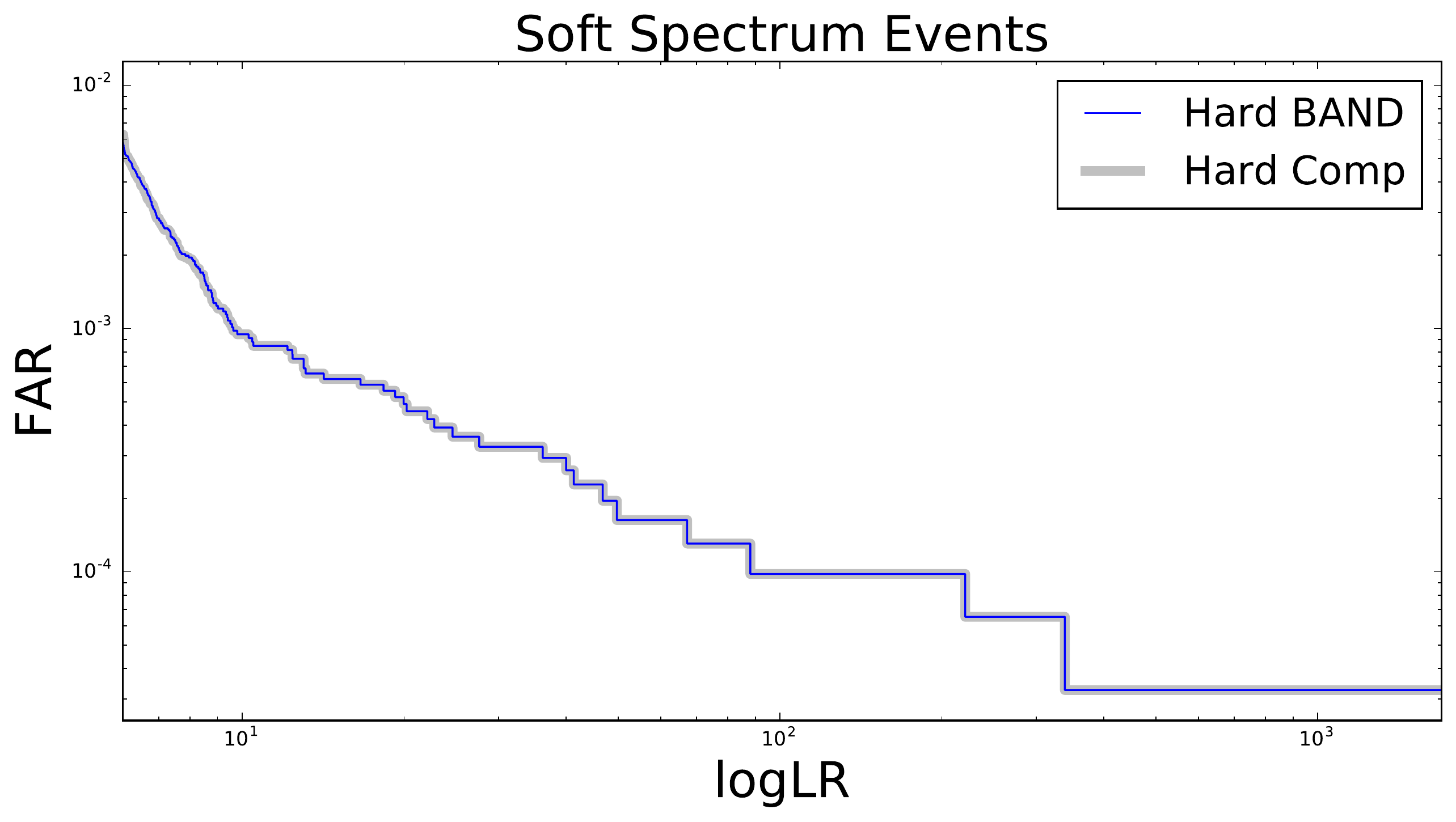}}
	\end{center}
\caption{The FAR comparison between the LB15 search using the `Hard' Band function template and the new `Hard' Comptonized 
template described in Section~\ref{sec:Templates}. The FAR distribution for the `Normal' template is affected by the change in the 
`Hard' template as some events are now found with a higher logLR in the `Hard' template.  The distribution for the `Soft' template 
is unaffected. \label{NewTempFAR}}
\end{figure}

In Figure~\ref{NewTempFAR}, we show the additional improvement made to the FAR by replacing the hard template in LB15 with 
the new Comptonized template.  Note that because the templates have some overlap in spectral parameter space, an event may 
be `detected' in more than one template, however, the search only considers the template with the highest ranking statistic.  The 
result of this can be seen in the difference between the FAR for the `Normal' spectral template when the `Hard' spectral template is 
replaced with the new Comptonized template:  some events that were `detected' with the `Normal' template were shifted over to 
the `Hard' template.  The `Soft' template is left unaffected by this change since it occupies a completely disjoint part of the 
parameter space relative to the `Hard' template.

\begin{figure}
	\begin{center}
		\subfigure[`Normal' Template]{\label{NewFinalMedium}\includegraphics[scale=0.26]{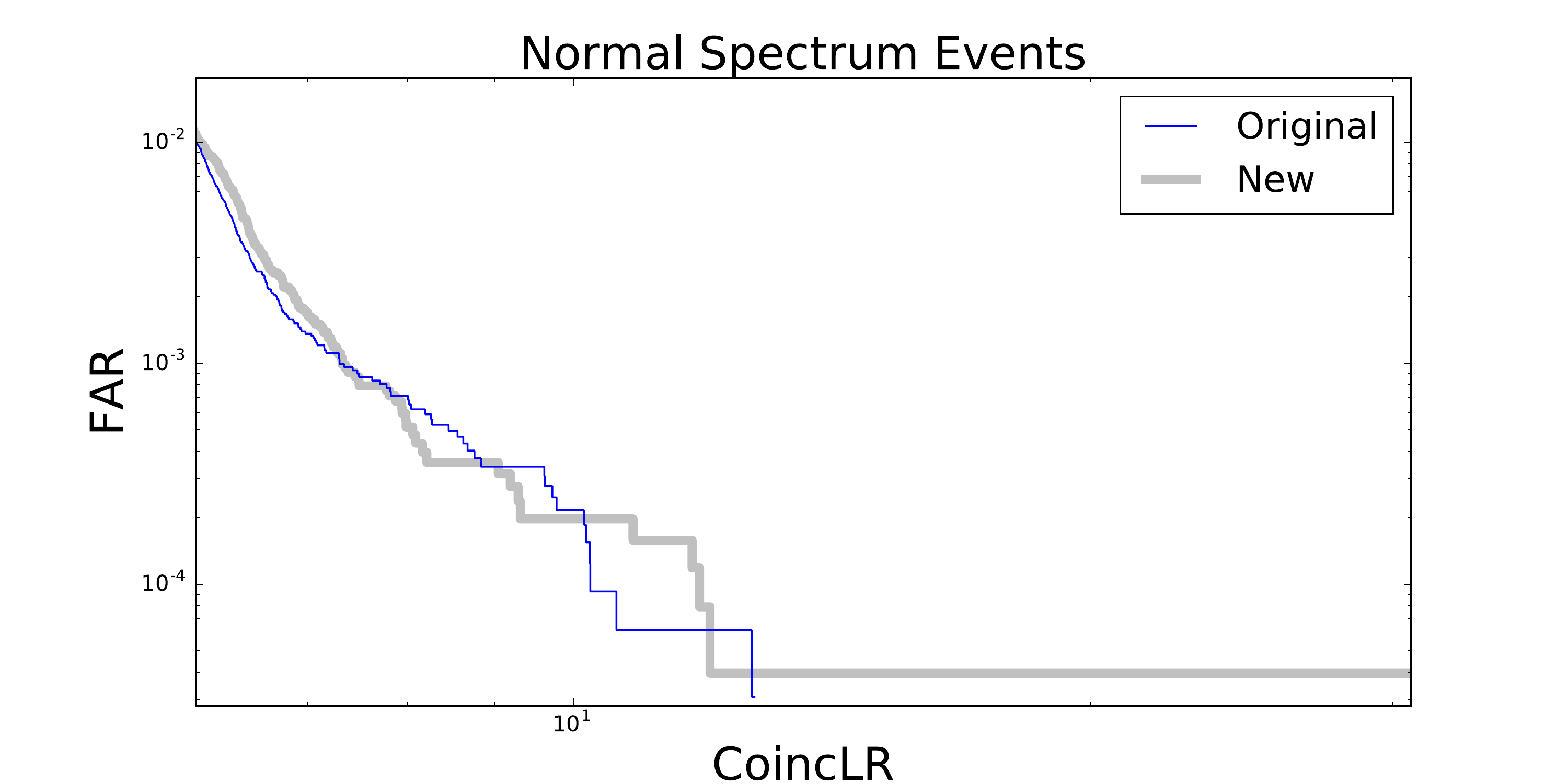}}
		\subfigure[`Hard' Template]{\label{NewFinalHard}\includegraphics[scale=0.26]{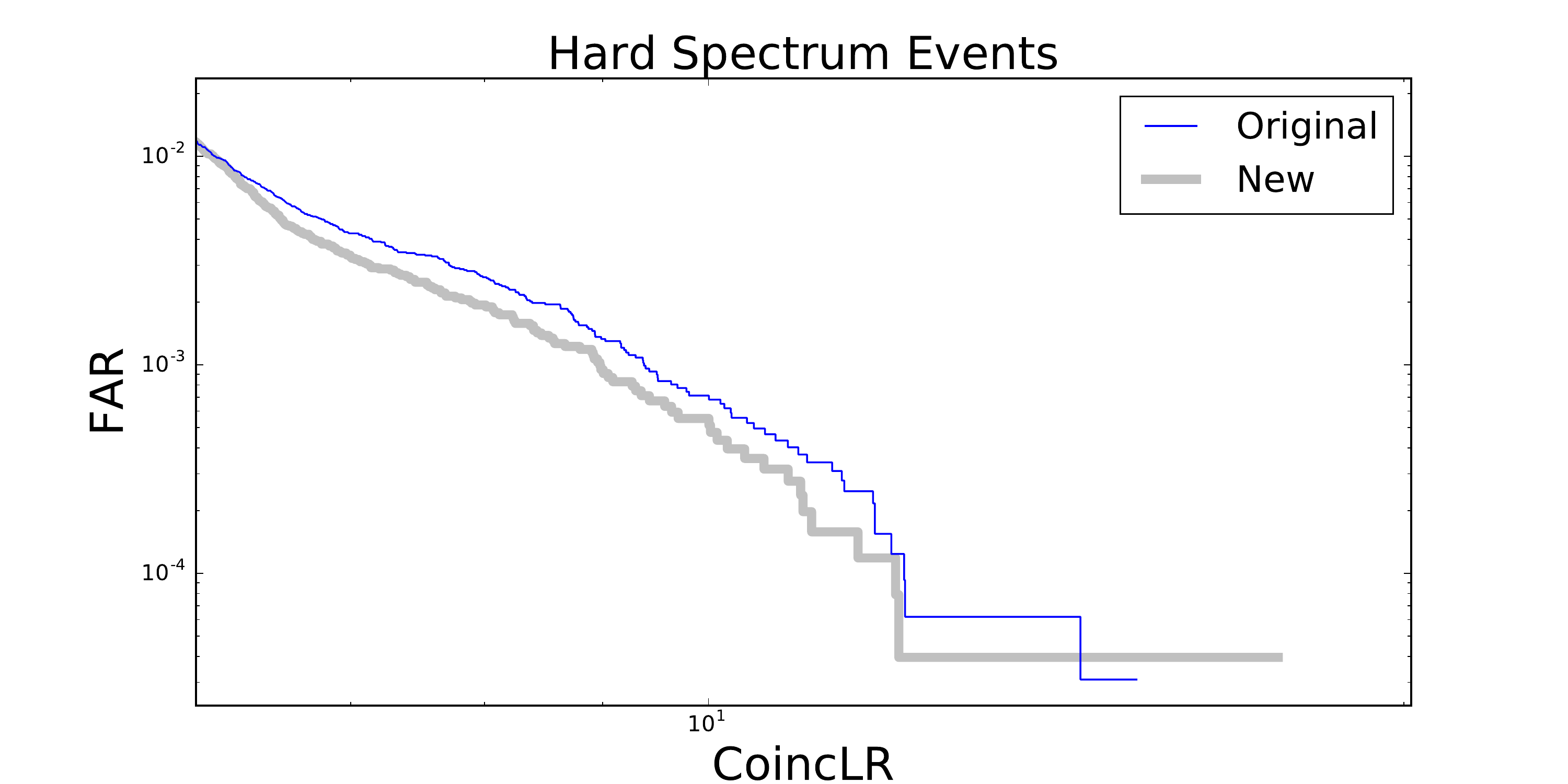}}\\
		\subfigure[`Soft' Template]{\label{NewFinalSoft}\includegraphics[scale=0.26]{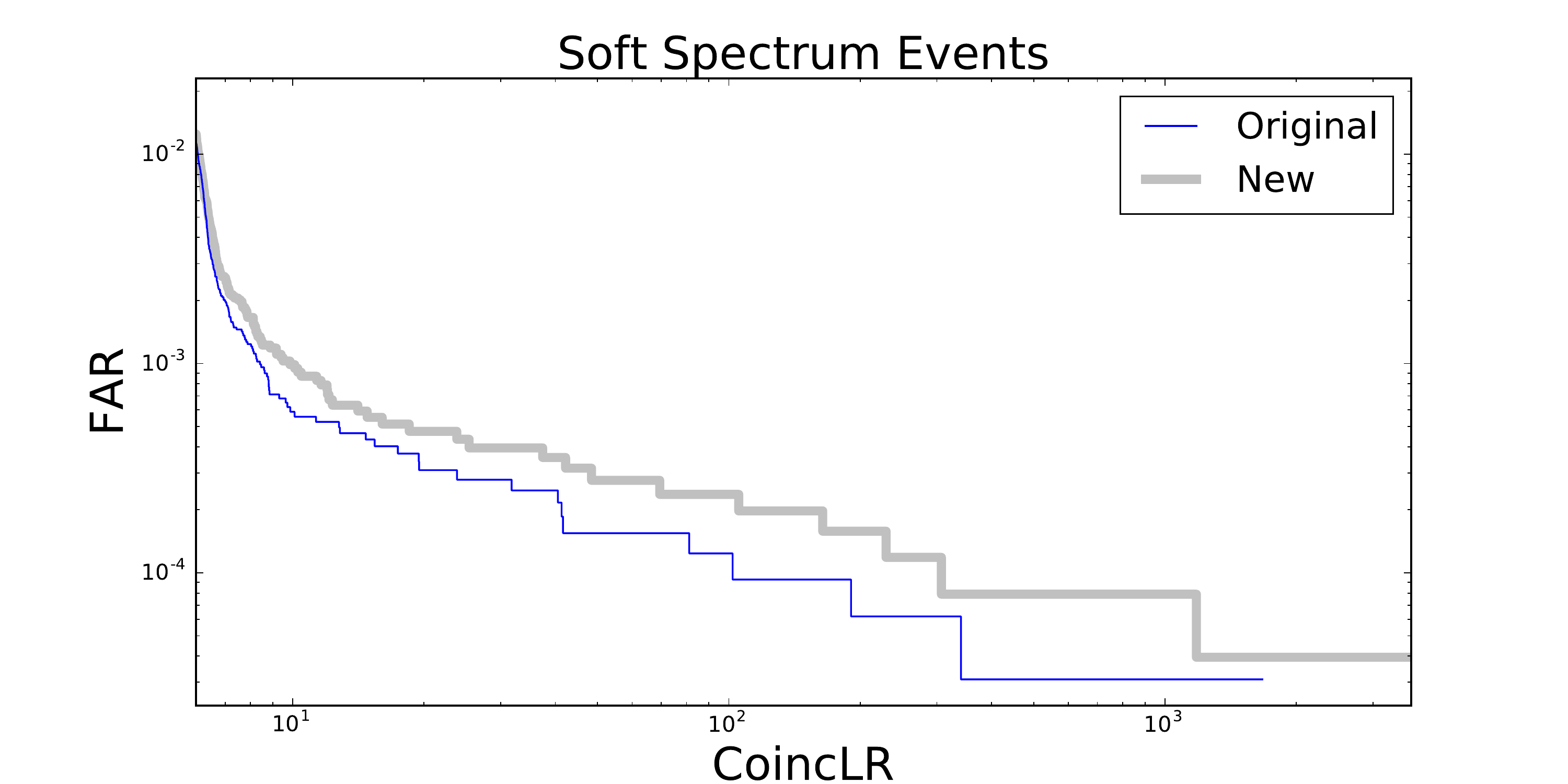}}
	\end{center}
\caption{The FAR comparison between the LB15 search (blue) and all of the combined updates (gray): new background, new 
`Hard' template, the spatial CoincLR ranking statistic, and the 64 ms binned data.
\label{NewFinalFAR}}
\end{figure}

Combining all of the aforementioned changes results in the final FAR comparison plots shown in Figure~\ref{NewFinalFAR}.  We 
find an obvious overall improvement in the FAR distribution for the `Normal' and `Hard' spectral templates compared to the LB15 
search.  The FAR distribution for the `Soft' spectral template does not show the same improvement.  This is likely due to 
the fact that GBM observes a large population of real signals, mostly Galactic, at lower energies, and the new background 
improves the detection of these events that are longer than 2 s.  Until a future improvement is implemented, such as a Galactic 
sky prior, any event of interest discovered by the `Soft' template will suffer a higher FAR and therefore a lower significance 
compared to the LB15 search.  The impact that the `Soft' template has on detecting a canonical short GRB, however, is minimal 
since almost all known short GRBs detected by GBM would be detected with the `Normal' or `Hard' template (possibly with the 
exception of precursors to the prompt emission).  Note that the use of the 64 ms binned data now allows more phase shifts for the 
256 ms and 512 ms timescales (up to a factor of four), thereby increasing the effective number bins searched.  This results in an 
overall higher FAR, however, the improvements from using the new background, the new `Hard' template, and the new ranking 
statistic more than offset this increase in the FAR.

%%Case Study
\section{Case Study: GW150914-GBM}
\begin{figure}
	\begin{center}
		\includegraphics[scale=0.5]{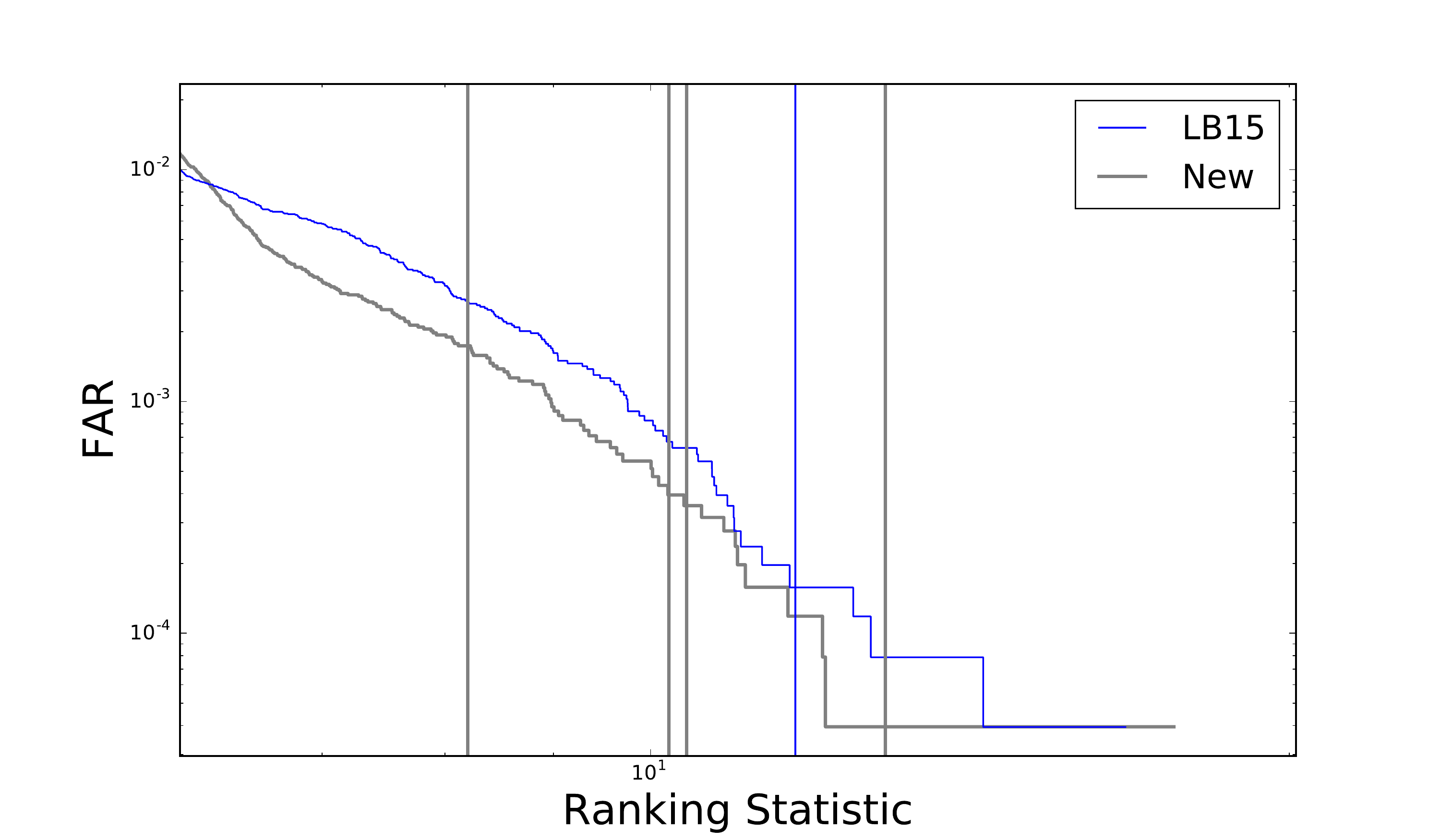}
	\end{center}
\caption{The comparison of the 150914-GBM candidate as detected by the LB15 search and the combined improvements for O2a.  
The blue line marks the LB15 ranking statistic value relative to FAR for the LB15 search, and the four gray lines mark the O2a 
ranking statistic relative to the O2a FAR for the Normal Phase and three 64 ms offsets relative to the LIGO detection time.
\label{150914FAR}}
\end{figure}

Although the updates to the LB15 search were not specifically motivated by the detection of a short, hard transient candidate 
$\sim$0.4 s after LIGO triggered on GW150914, the presence of such a candidate allows us to compare how the updated 
search fares relative to the LB15 search that found the candidate.  Figure~\ref{150914FAR} shows this comparison with the 
ranking statistic of the candidate marked on both FAR distributions.  The FAR distribution using the LB15 search covers the same 
time as the FAR distribution using the O2a search for a fair comparison and is therefore slightly different than the distribution found 
in~\citet{Connaughton16}.

The LB15 search found that the most significant signal was in the 1.024 s timescale, and because the CTIME datatype is natively 
binned at 256 ms resolution, this allows a phase shift factor of 4 for the 1.024 s timescale.  The search using the TTE data binned 
at 64 ms allows a phase shift factor of 4 for the 256 ms timescale, however we have maintained the LB15 maximum phase shift 
factor of 4 for any single timescale, which means that we still only perform a factor 4 phase shift for the 1.024 s timescale, as 
described in Section~\ref{sec:InputData}.  This implies that the O2a search will not necessarily use the same binning phase used 
in the O1 search.  In fact, Figure~\ref{150914FAR} shows the result of maintaining the factor 4 phase shift for the 1.024 s 
timescale by offsetting the search window by 64 ms, four times.  The ranking statistic varies from 8.2--12.9 (the largest value 
corresponding to the phase closest to the O1 search phase), which results in a pre-trials FAR ranging from 
$1.7\times10^{-3}$--$3.9\times10^{-5}$ Hz (the O1 pre-trials FAR was $1.2\times10^{-4}$~Hz).  The maximum relative change in 
the ranking statistic for the 64 ms phase shifts is 36\%, consistent with what was found in Section~\ref{sec:SignalValidation}.

%%References

\clearpage

%%Figure 1

%%Figure 2

%%Figure 3

%%Figure 4

%%Figure 5

%%Figure 6

%%Figure 7

%%Figure 8

%%Figure 9

%%Figure 10

%%Figure 11

%%Figure 12

%%Figure 13

\end{document}